\documentclass[preprint,12pt,3p]{elsarticle}

\usepackage[utf8]{inputenc}
\usepackage[english]{babel}

\usepackage[fleqn]{amsmath}
\usepackage{amssymb}
\usepackage{cases}

\usepackage{soul, xcolor}

\usepackage{algorithmic, algorithm}

\usepackage{multicol}

\journal{Simulation Modelling Practice and Theory}

\begin{document}

\begin{frontmatter}

\title{A Family of Simulation Criteria to Guide DEVS Models Validation Rigorously, Systematically and Semi-Automatically}

\author[cif]{Diego A. Hollmann\corref{DAH}}
\ead{hollmann@cifasis-conicet.gov.ar}

\author[cif,unr]{Maximiliano Cristi\'a}
\ead{cristia@cifasis-conicet.gov.ar}

\author[cif,amu]{Claudia Frydman}
\ead{claudia.frydman@lsis.org}

\cortext[DAH]{Corresponding author}
\address[cif]{CIFASIS - CONICET.\\ Centro Internacional Franco Argentino de Ciencias de la Informaci\'on y de Sistemas.\\27 de Febrero 210 Bis, S2000EZP Rosario, Argentina.}
\address[unr]{FCEIA - UNR, Rosario, Argentina.}
\address[amu]{LSIS - AMU, Marseille, France.}

\begin{abstract}
The most common method to validate a DEVS model against the requirements is to simulate it several times under different conditions, with some simulation tool. The behavior of the model is compared with what the system is supposed to do. The number of different scenarios to simulate is usually infinite, therefore, selecting them becomes  a crucial task. This selection, actually, is made following the experience or intuition of an engineer. Here we present a family of criteria to conduct DEVS model simulations in a disciplined way and covering the most significant simulations to increase the confidence on the model. This is achieved by analyzing the mathematical representation of the DEVS model and, thus, part of the validation process can be automatized.
\end{abstract}


\end{frontmatter}

\section{Introduction}
\label{Intro}

The development and use of simulation models has been increasing considerably in recent years. Frequently they are used as the first representation of systems that later will be used for decision-making on critical situations. It has become increasingly necessary the definition of rigorous techniques to assure that these models represent as well as possible the real system being modeled. In other words, Verification and Validation (V\&V) of simulation models has become crucial.

According to the U.S. Department of Defense directive \cite{DoDD94} verification is ``the process of determining that a model implementation accurately represents the developer's conceptual description and specifications''. In the context of simulation models, the question that model verification tries to answer is: ``Are we simulating, or have we simulated, the model right?''. On the other hand, validation is ``the process of determining the degree to which a model is an accurate representation of the real-world from the perspective of the intended uses of the model''. Here, the question is: ``Are we simulating, or have we simulated, the right model?''.

Performing V\&V of simulation models has been identified as a paramount activity as it can increase the confidence of the user in the simulation results and lead to the accreditation/certification of the simulated system \cite{LW05}. This accreditation or certification is specially important when the simulation results influence decision making over crucial issues.

\begin{figure}[h]
\centering
\includegraphics{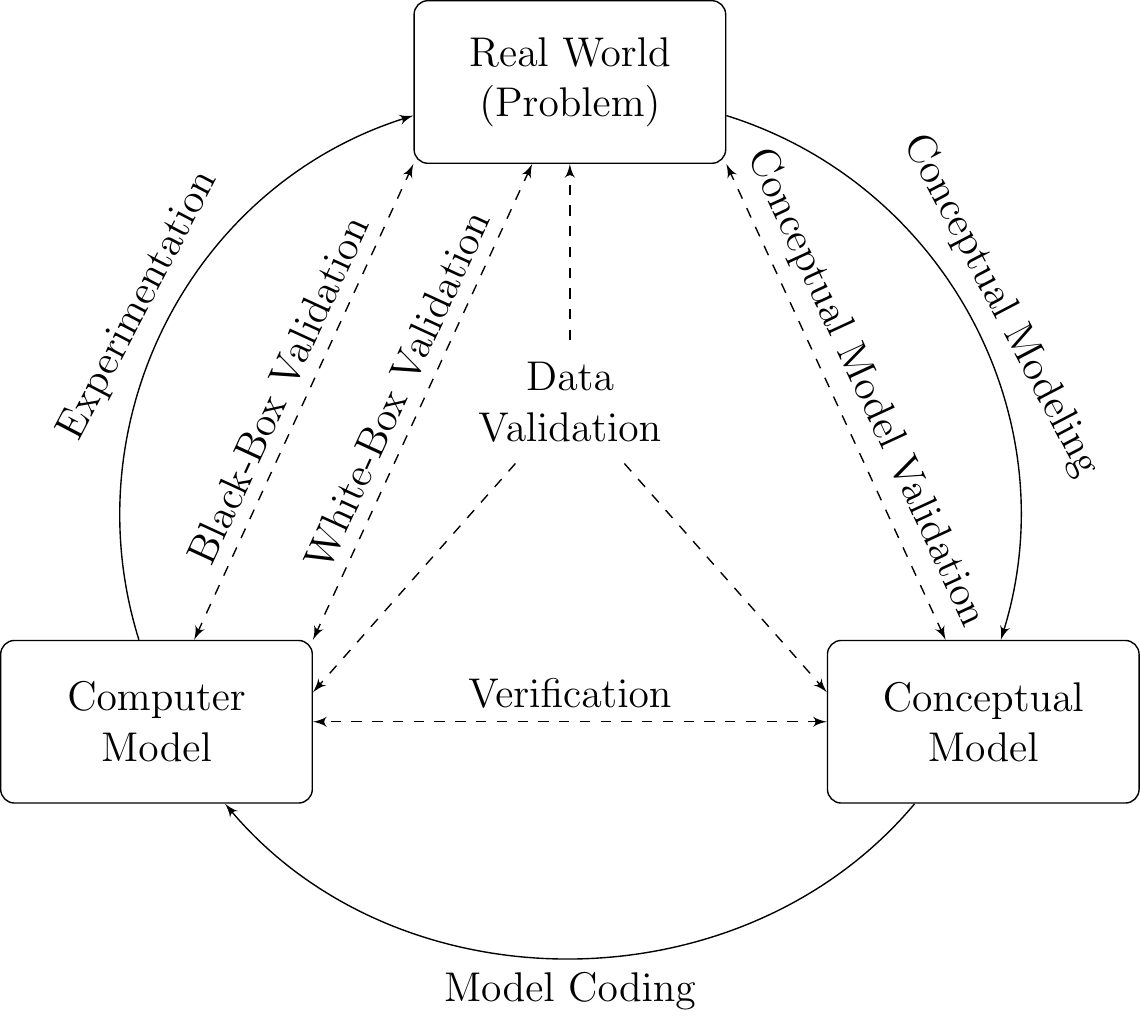}
\caption{Simulation Model Verification and Validation in the Modeling Process}
\label{ModProc}
\end{figure}

Figure \ref{ModProc}, presented by Robinson \cite{Robinson97}, which, in turn, is adapted from Sargent \cite{Sargent92} shows the different phases of V\&V onto the modeling process. Our work is related to the \textit{Conceptual Model Validation} phase, i.e., validate the conceptual or abstract simulation model against the requirements.

\subsection{Validation of Simulation Models and Model-Based Testing}
\label{ValDEVS}
The validation of a model against the requirements, usually, cannot be performed mathematically because the requirements are not formal. An alternative way is via simulations. The engineer compares the results of the simulations with the requirements in order to decide if the model is correct or not. This is particularly important when either the model is large, its implementation is critical or critical decisions are made according to the results provided by the implementation of the model or by its simulations. Besides, since it is very important to find as many errors as possible and as earlier as possible, then a thorough simulation process can be an activity that will reduce the total cost of ownership of the target system.

During model simulation it would be desirable to run simulations from all possible simulation configurations and compare these behaviors against the requirements. Unfortunately, exhaustive simulation is impractical in almost all projects, since it involves an infinite number of simulation configurations. Considering this, the selection of an appropriate set of simulation configurations is a crucial issue that should consider two opposite factors: a) the set of simulation configurations should be large enough as to give a reasonable assurance that the model is a correct representation of the requirements; and b) the set should be small enough so V\&V fits within time and budget.

In this paper DEVS (Discrete EVent system Specification) \cite{ZPK00} is used as the modeling formalism. DEVS has gained popularity in recent years and it is the most general formalism to describe discrete event systems (DES). DEVS is an abstract basis for model specification that is independent of any particular simulation implementation. It is a formalism based on system theory, expressive enough to represent all other DES formalisms, i.e. all models representable in those formalisms can be represented in DEVS \cite{ZV93}.

\begin{figure}[ht]
\centering
\includegraphics{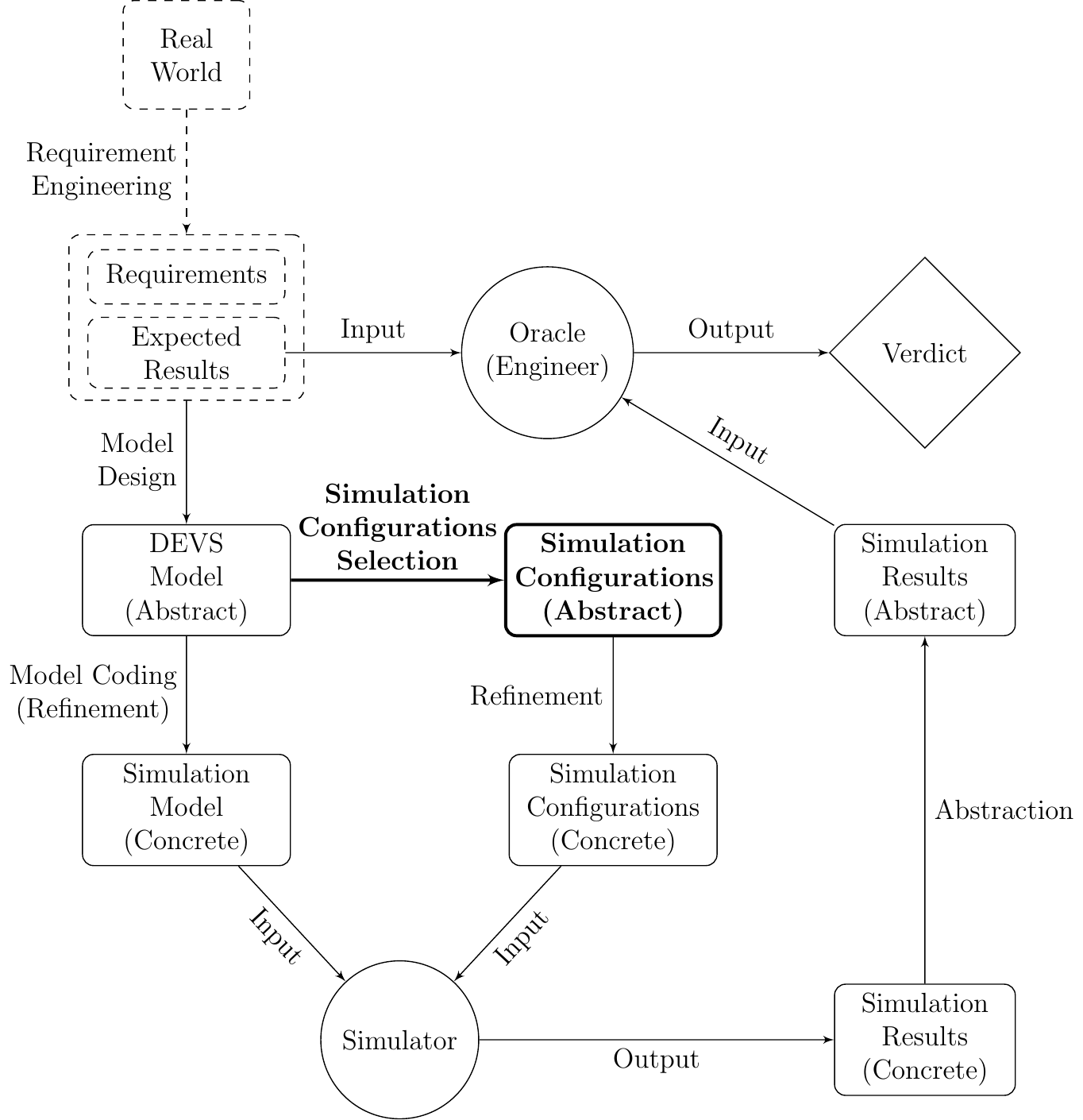}
\caption{DEVS Model Validation via simulations}
\label{DEVS_Validation}
\end{figure}

Figure \ref{DEVS_Validation} presents a possible DEVS model validation process via simulations. According to this picture, first of all, the requirements and expected results are extracted from the \textit{Real World}, or from the depiction of the problem being modeled. Based on them, the conceptual or abstract model is defined using a modeling formalism, in this case, DEVS. At this point, it is important to point out that, in order to simplify the validation process, this model must be described using a mathematical or logical representation of DEVS. 

According to this validation process, once the abstract DEVS model is defined, a set of simulations configuration is derived from it. Afterwards, both, the simulations and the abstract DEVS model are refined, i.e written in the input language of some concrete simulator (for instance, DEVS-C++ \cite{CC97}, DEVSim++ \cite{Kim94}, CD++ \cite{Wainer02}, PowerDEVS \cite{BK10}, JDEVS \cite{FDB02}). That is, in this context refinement means to translate an abstract representation into a concrete one. The concept of refinement is borrowed from the software engineering community \cite{UL06}

Finally, the (concrete) simulation results need to be abstracted. In this sense, abstraction is the inverse process of refinement. Later, these abstract results are compared with the expected results. This latter issue, generally known as the \textit{oracle problem}, is acute in this context as it is, for instance, in software testing, and there is no general solution \cite{LW05,Hetal09}. 

Despite the fact that the final goal is to formalize the whole validation process, this article is focused on the selection of an appropriate set of simulation configurations. As mentioned before, this is the crucial part of this validation process. It consists in turning an infinite problem (the set of all simulations configurations) into a finite one. Further, this must be done trying to keep in the final set the important or revealing simulations. In order words, the final set of simulation configurations must be small and must thoroughly cover all the functional alternatives described in the model. The rest of the validation process consists mainly in an implementation process and is further detailed in Section \ref{Auto}.

In general, model simulation is performed according to the experience or intuition of a specialist. Therefore no rigorous guidelines or criteria are followed to define an adequate set of simulations, making the simulation process informal and error prone. On the other hand, given that the selection of the simulations is an informal activity it cannot be automated in a high degree. However, it can be automated to some extent if the selection process is formalized in such a way that, later, a software tool can help in this task. Therefore, it would be desirable to formally define simulation criteria to consider the simulation of a model as a validation process with an acceptable degree of accuracy.

In the software testing field there is an analogous scenario. According to Utting and Legeard \cite{UL06} software testing deals with the dynamic verification of the behavior of a program on a finite set of test cases, suitably selected from the usually infinite execution domain, against the expected behavior. There are several works trying to formalize the software testing process. Most of these works belong to a subfield of testing known as Model-Based Testing (MBT). Utting and Legeard define MBT as the generation of executable test cases, based on models of the behavior of the system under test (SUT). Hence, an analogy can be established between MBT and model validation by simulation. The model can be seen as the specification of the SUT in MBT and the test case generation as the generation of simulation configurations.

Being so important in software development, the MBT process has been improved up to the point of turning it almost automatic, in many cases obtaining quite good results \cite{Hetal09,UL06,CristiaSTVR}. Thus, it is worth to explore if MBT techniques can be replicated in the context of model validation by simulation.

An important part of this automation is possible because there are precise testing criteria. That is, testing criteria indicate which tests must be generated from the model, since there is usually an infinite number of possible tests \cite{UL06}. Some of these criteria are based on the exploration of the model and others on the exploration of the source code of the program. 

Even though many MBT methods can be used or adapted to follow the analogy mentioned before, we based our work on the Test Template Framework (TTF). The TTF is a MBT method presented by Stocks and Carrington \cite{SC96} and implemented by Cristi\'a et al. \cite{CristiaSTVR}. The TTF was introduced to formally define test data sets providing structure to the testing process. The selection of the TTF as a MBT method is motivated by the fact that it deals with the logical and mathematical definition of the model instead of analyze, for instance, traces or executions of the model. Therefore, it can be applied to systems more complex than Finite State Machines (FSM) and DEVS is much more general than a simple FSM.

Taking all of this into account, the idea of this work is to extend the rigorous validation process of DEVS models presented by Hollmann et al. \cite{HCF12}. Also, a possible way to automate this process is discussed here. Furthermore, a new case study is presented and, in addition, we show some possible errors that could be overlooked during the validation or simulation phase if it is not carried following some systematic and disciplined criteria. The major contribution of this paper is to present an alternative systematic and semi-automatic method to conduct the simulation process of DEVS models in order to validate them. This alternative method is based on well-known techniques in the \textit{testing world} that can be adapted for the \textit{model and simulation community}. We believe that simulating DEVS models following the techniques presented in this work will increase the confidence that the model is correct validating aspects or features of the model that could be overlooked by the specialist.

It is important to point out that the issue of running the simulations is not faced in this paper, leaving it as a second step of our research. Here, we give formal criteria to generate the simulation configurations that allow the corresponding simulations to be run in order to validate the model. However, we discuss how this work can be extended in such a way that simulation configurations can be provided to simulation tools.

The remainder of this paper is organized as follows. In the following section we describe and comment some other approaches to model validation. Section \ref{SimCrit} presents the simulation criteria that form the core of our contribution. The possible automation of this validation process is addressed in Section \ref{Auto}. In Section \ref{CaseStudies} two case studies are described, and in section \ref{Concl} conclusion and some future work is discussed.

\section{Related Works and Similar Approaches}
\label{RelWork}
Below we discuss and comment different works involving verification and validation of simulation models. These works either present the most similar approaches to the one presented here, or they motivate or show why our work should be relevant for the modeling and simulation community.

Balci \cite{Balci97} presents guidelines for conducting verification, validation and accreditation of simulation models. He made a classification of the different V\&V techniques for simulation models presenting a taxonomy of more than 77 V\&V techniques for conventional simulation models and 38 V\&V techniques for object-oriented simulation models. In his work, Balci listed model testing as one of the candidates for V\&V of simulation models. Also, Labiche and Wainer \cite{LW05} make a review of the V\&V of discrete event system models. They propose to apply or adapt existing software testing techniques to the V\&V of DEVS models. In particular, they claim that formal techniques should be applied. Thus justifying our approach. In several works \cite{Sargent03,Sargent05,Sargent07,Sargent10} Sargent discusses different approaches, paradigms and techniques related to validation and verification of simulation models. These are interesting works to learn about a generalization of different validation and verification processes, however they do not describe any particular validation process in detail. Our present work complements all these papers by giving a detailed, semi-formal validation method adapted from the software engineering community.

There are several works that use verification techniques, like model checking, to verify the correctness of a model. For instance, Napoli and Parente \cite{NP11} present a model-checking algorithm for Hierarchical Finite State Machines as an abstract DEVS model. They also focus on the generation of simulation configurations for DEVS, but as counter-examples obtained by the application of their model-checking algorithm. Another relevant and recent work involving verification techniques is \cite{SW13} where Saadawi and Wainer introduce a new extension to the DEVS formalism, called the Rational Time-Advance DEVS (RTA-DEVS). RTA-DEVS models can be formally checked with standard model-checking algorithm and tools. Further, they introduce a methodology to transform classic DEVS models to RTA-DEVS models, allowing formal verification of classic DEVS. Although model checking techniques are formally defined and they are useful to prove properties and theorems over a model, the main problem of such techniques is the so-called \textit{state explosion problem} \cite{BK08}, i.e. the exponential blowup of the state space and variables in any real or practical system. This made almost impossible the use of such techniques in large projects, although model checking has been used in real projects.

K. J. Hong and T. G. Kim \cite{HK05} introduce a method for the verification of discrete event models. They propose a formalism, Time State Reachability Graph (TSRG), to specify modules of a discrete event model and a methodology for the generation of test sequences to test such modules at an I/O level. Later, a graph theoretical analysis of TSGR generates all possible timed I/O sequences from which a test set of timed I/O sequences with 100\% coverage can be constructed. Similar to our work, they make an analogy between model verification and software testing. 

Another recent work that applies verification techniques over discrete event simulation is \cite{dSdM11} where da Silva and de Melo presents a method to perform simulations orderly and verify properties about them using transitions systems. Both, the possible simulation paths and the property to be verified are described as transition systems. The verification is achieved by building a special kind of synchronous product between these two transition systems. They focused their work on the verification of properties by simulation but not on the generation of simulations in order to validate the model.

Li et al. \cite{Letal11} present a framework to test DEVS tools. In their framework they combine black-box and white-box testing approaches. Actually, this work is not really related to ours because they do not validate or verify a DEVS model, whereas they test DEVS implementations. However, it is useful to see how they introduce software testing techniques in the DEVS world.

After reviewing many works about V\&V for simulation models it seems that there is no proposals similar to the validation method presented in this article. For instance, we could not find any work that deals with the mathematical or logical representation of the model as a starting point of the validation process. Moreover, we think that, in general, people of the modeling and simulation community do not take this issue into account. Actually they develop their models directly over a simulation tool, using its own modeling language, and make experiments directly over it. By working with the concrete model, these techniques aim at a verification problem, i.e. comparing the concrete model with the abstract model. We, on the other hand, propose, as a complementary technique, to work with the abstract model to check whether it conforms with the user requirements or objectives. More concretely, our work proposes a method to formally validate abstract models and to introduce well known MBT techniques into this community.

\section{A family of simulation criteria}
\label{SimCrit}
The simulation criteria presented in this work aim to provide a guideline for methodically simulate DEVS models in order to validate them. As we mentioned in the introduction, DEVS is a formalism widely used in the modeling and simulation community and is the most general formalism to describe discrete event systems.

According to Zeigler et al \cite{ZPK00}, there are two classes of models, \textit{Atomic} models and \textit{Coupled} models. Our work concerns only atomic models. In fact, a coupled model is equivalent to a (complex) atomic model \cite{ZPK00}.

An atomic DEVS Model is defined by the structure:
\begin{center}
$M = (X, Y, S, \delta_{int}, \delta_{ext}, \lambda, ta)$
\end{center}
where:
\begin{itemize}
    \item $X$ is the set of input event values, i.e., the set of all possible values that an input event can assume;
    \item $Y$ is the set of output event values;
    \item $S$ is the set of state values;
    \item $\delta_{int}: S \rightarrow S$ is the internal transition function;
    \item $\delta_{ext}: Q \times S \rightarrow S$ is the external transition function, where:
        
        \hspace{0.5cm} $Q=\{(s,e), s \in S, 0 \leq e \leq ta(s)\}$ is the \textit{total state} set, and
        
        \hspace{0.5cm} $e$ is the \textit{time elapsed} since last transition;
    \item $\lambda: S \rightarrow Y $ is the output function; and
    \item $ta: S \rightarrow \mathbb{R}^+_{0,\infty}$ is the time advance function.
\end{itemize}
$\delta_{int}$, $\delta_{ext}$, $\lambda$ and $ta$ are the functions that define the system dynamics. Each possible state $s \in S$ has associated a time advance, $ta(s) \in \mathbb{R}^+_{0,\infty}$, which indicates the time that the system will remain in that state if no input events occur. Once that time has passed, an internal transition is performed, reaching a new state $s'$, $s'=\delta_{int}(s)$. At the same time, an output event, $y$, is generated by the output function, $y=\lambda(s)$. Therefore, $\delta_{int}$, $ta$ and $\lambda$ define the autonomous behavior of the system.

When an input event arrives, an external transition is performed. The new state depends on the input value, the previous state and also the elapsed time since the last transition. If the system is in the state $s$ and the input event $x$ arrives in the instant $e$ (i.e. $e$ time unites the last transition) the new state, $s'$, is calculated as $s'=\delta_{ext}(s,e,x)$. In case of an external transition, no output event is generated.

\subsection{Simulation Configurations Partition}
\begin{figure}[h]
\centering
\includegraphics{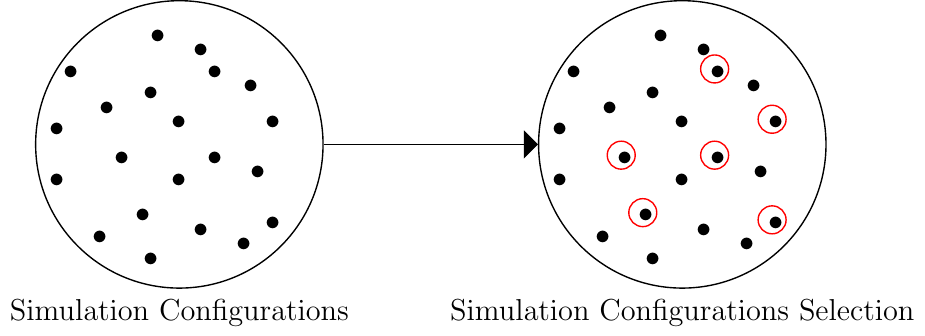}
\caption{Traditional Configurations Selection}
\label{TradSimSel}
\end{figure}

Given a DEVS model, in order to simulate it with some simulation tool, it is necessary to provide an initial state (not defined by the formalism) and a sequence of input events with their corresponding occurrence time, we call this initial state and sequence of input event a \textit{Simulation Configuration}. Usually, the set of possible simulation configurations (all possible initial state and all possible input events) is infinite, even if the model has finite sets of inputs, states and outputs. Therefore, validation via simulation requires to select some of these possible simulation configurations, see Figure \ref{TradSimSel}. Currently, this selection is done according the experience of the engineer, therefore, a domain expert is usually needed.

\begin{figure}[ht]
\centering
\includegraphics{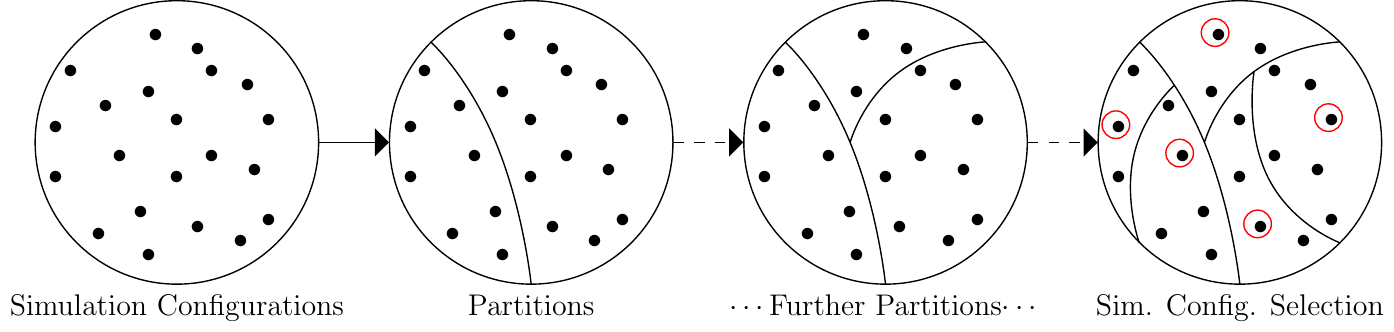}
\caption{Our Proposal: Simulation Configurations Partition}
\label{SimConfigPart}
\end{figure}

The technique presented in this work proposes to divide the set of all possible simulation configurations into equivalence classes by applying one or more \textit{partition criteria}. We call each of these equivalence classes a \textit{Simulation Configuration Class} (SCC), i.e. a SCC is a set of possible simulation configurations. Afterward, one simulation configuration of each SCC must be selected, see Figure \ref{SimConfigPart}. These selected simulation configurations are the only one that should be executed. 

The reason for selecting just one simulation configuration from each class is based on the \textit{uniformity hypothesis} presented by Bougé et al. \cite{BCFG86}. They assert, in software testing, that ``A program behaves uniformly on a equivalence class if the following holds: if the program works correctly for some input data of the equivalence class then it works correctly for any of them.'' This is a key assumption made by the software testing community because it allows to reduce the potentially infinite input domain to a small, finite one. Being an assumption it is not proved although many testing methods are based on it \cite{Hetal09}. In fact, the very idea of testing is implicitly based on this hypothesis because a single test case actually represents a whole class of test cases. In other words, when a tester selects a test case, he or she is assuming that it represents a set of possible candidates.

We adapt this concept to model validation. Therefore, we say that these subsets, in which the set of possible simulations was divided, are equivalence classes because it is assumed that the model has an uniform behavior for each subset of simulations. Furthermore, if the uniformity hypothesis holds and an error in the model is found for some  simulation of a given class, then the same error would be revealed with any other simulation of that class.

In this context, the uniformity hypothesis can be formalized as follows. Let $M_{abs}$ be some abstract model ($M_{abs}$ can be seen as a mathematical or logical formula), $M_{con}$ its concrete model, i.e. its corresponding refinement in a simulation tool language. If $a$ is a simulation configuration derived from $M_{abs}$, let $a'$ be its refinement. Then, $M_{con}(a')$ means the execution of $a'$ on $M_{con}$; and $M_{abs}(a,M_{con}(a'))$ asserts that $M_{con}(a')$ is the expected result with respect to $a$ according to $M_{abs}$. Note that if $M_{con}(a')$ is not the expected result of $a$ according to $M_{abs}$, then $M_{abs}(a,M_{con}(a'))$ if false (i.e. $\neg M_{abs}(a,M_{con}(a'))$ is true). Then, the uniformity hypothesis holds if and only if for every $SCC_i$ and $a \in SCC_i$ the following holds:
\begin{center}
$M_{abs}(a,M_{con}(a')) \Rightarrow \forall\ x \in SCC_i: M_{abs}(x,M_{con}(x'))$
\end{center}
that is, the uniformity hypothesis holds if the model behaves the same for any two elements of a given $SCC_i$.

With this uniformity hypothesis, some strategies or criteria assume that every element of a class is equivalent to all the others (of the same class) for the simulation of a particular functionality of the model. However, this is just an hypothesis and it is not always satisfied. Therefore, as noted by Stocks and Carrington \cite{SC96}, this assumption is often invalid and they propose to apply repeatedly the strategies in order to partition the classes into sub-classes until either the engineer considers that the classes are reasonable small or each functionality of the model is covered by only one class \cite{CM09}.

Each element of a SCC consists of an initial state (to initialize the simulation) and an input pair containing the event to simulate and its corresponding time, i.e. when the event must be simulated.

Following this idea, in order to describe a SCC we need to define the set of possible initial states and the set of possible input pairs for the simulation. An input pair is an ordered pair $(event, time)$. Therefore a SCC is defined by:
\begin{itemize}
 \item a set of states, $IniSt \subseteq S$, and
 \item a set of ordered pairs of event and time, $InPairs \subseteq  \{(x, t): x \in X \cup \{\tau\}, t \in \mathbb{R}^+_0\}$.
\end{itemize}
where $\tau$ represents the \textit{no event} situation and an internal transitions occur in that case. This is necessary to indicate the simulation of an internal transition.

It is important to point out that in the DEVS formalism no initial state is defined. This, actually, belongs to the simulation phase and, since this work aims to conduct these simulations, an initial state must be defined. This is not necessarily the initial state of the system it is just the starting state for the simulation. Moreover, defining the initial state makes the simulation of the different functionalities of the model simpler, since it would not be necessary to guide the system to a particular state and start a simulation from there.

The total class of simulations, i.e. the set of all possible simulations, for a given DEVS model is defined by:
\begin{itemize}
 \item $IniSt = IniSt_1 \cup \ldots \cup IniSt_n = S$, and
 \item $InPairs = InPairs_1 \cup \ldots \cup InPairs_n = \{(x,t): x \in X \cup \{\tau\} \wedge t \in \mathbb{R}_0^+)\}$.
\end{itemize}
$IniSt$ represents all possible initial states from which a simulation can be started. Note that, initially, all the states of the model, i.e. $S$, are potential candidates. $InPairs$ is the set of all possible input pairs. Therefore, the idea is to partition $IniSt \times InPairs$ by analyzing the DEVS model guided by criteria defined below.

\subsection{Partition Criteria}
\label{PartCrit}
Now we present the different partition criteria proposed in this work. The criteria are applied to different aspects of DEVS models. Some criteria apply, for instance, to the external transition function definition, others to the internal transition function definition and others to the definition of the states or input and outputs sets.

It is important to mention that at any time of the partition process it is possible that some criteria could not generate new classes, this is because the result could be classes already obtained. Moreover, it does not matter the order in which the criteria are applied, since each criterion is independent from the others. 

In each of the following subsections a partition criterion is described.

\subsubsection{Transition Functions Defined by Cases}
It is very common to define the external and internal transition functions by cases. The first and more intuitive criterion is to partition the set of possible simulations into several classes, one for each case in the definition of the function. 
    
Let $\delta_{ext}$ and $\delta_{int}$ be the transition functions of a DEVS model defined by cases:
\begin{center}
$\begin{aligned}
\delta_{ext}(s,e,x)&=
\begin{cases}
    expr_{ext}^1(s,e,x) \text{ if } P_{ext}^1(s,e,x) \\
    \vdots \\
    expr_{ext}^n(s,e,x) \text{ if } P_{ext}^n(s,e,x) \\
\end{cases} \\
\delta_{int}(s)&=
\begin{cases}
    expr_{int}^1(s) \text{ if } P_{int}^1(s) \\
    \vdots \\
    expr_{int}^m(s) \text{ if } P_{int}^m(s) \\
\end{cases}
\end{aligned}$
\end{center}
where $expr_{ext}^i$ and $expr_{int}^j$ are the results of the function if the proposition $P_{ext}^i$ or $P_{int}^j$, respectively, holds. This criterion proposes to generate one class for each proposition in the definition of the internal and external transition functions. Each class is defined by:
\begin{itemize}
\item Associated to the External Transition Function:
\begin{center}
$\begin{aligned}
IniSt_i &= \{s \in S \mid \exists\ e \in \mathbb{R}_0^+, x \in X : P^i_{ext}(s,e,x)\}, \\
InPairs_i &= \{(x,t) \in InPairs \mid \exists\ s \in IniSt_i, e \in \mathbb{R}_0^+ : P^i_{ext}(s,e,x)\}.
\end{aligned}$
\end{center}
with $i \in [1, n]$
\item Associated to the Internal Transition Function:
\begin{center}
$\begin{aligned}
IniSt_j &= \{s \in S \mid P^j_{int}(s)\}, \\
InPairs_j &= \{(\tau, 0)\}.
\end{aligned}$
\end{center}
with $j \in [1, m]$.
\end{itemize}

In the case of the internal transition the idea is to configure a certain state allowing a particular internal transition to occur, that's why the time relative to the $\tau$ event is 0.

\subsubsection{Extensional Sets}
In the definition of DEVS models, sometimes, some sets are defined by extension (states, input events or output events), i.e. listing the elements of the set. Necessarily these sets will be finite and relatively small. Therefore, this criterion proposes to simulate all scenarios where appears at least once each element of these sets.

Let us suppose that the set of states values, $S$, of some DEVS model is an extensional set:
\begin{center}
$S = \{s_1, s_2, \ldots , s_n\}$
\end{center}
therefore, the classes generated by applying this criterion should be:
\begin{center}
$\begin{aligned}
IniSt_i &= \{s_i\},\\
InPairs_i &= InPairs.
\end{aligned}$
\end{center}
with $i \in [1, n]$. By applying this criterion the engineer guarantees the simulation of the model in each possible state. However, if this criterion is only applied over the state definition, and no other criterion is applied, certain input events would not be simulated for certain states.

Let us suppose now, that the set of input variables is defined as:
\begin{center}
$X=\{x_1, x_2, \ldots, x_n\}$
\end{center}
the resulting classes now would be:
\begin{center}
$\begin{aligned}
IniSt_i &= S,\\
InPairs_i &= \{(x_i,t), t \in \mathbb{R}^+_0\}.
\end{aligned}$
\end{center}
with $i \in [1, n]$. Herein, the engineer ensures to simulate all possible input events. Combining these classes with the former, allows the simulation of all states with all inputs. This is explained further in section \ref{CombPart}.

\subsubsection{Intentional Sets}
Set comprehensions or intentional sets is another usual form for defining sets in a DEVS model, i.e. specifying the properties that each element of the set must comply. This is done by a logical predicate, that could be a simple one or a very complex definition involving several operations. Thus, this criterion proposes to use this definition to partition the set of simulation configurations.

Let us suppose that the set of state is an intentional set, $S=\{s: TYPE \mid P(s)\}$, where $TYPE$ is the type of the elements of the set and $P$ is a logical predicate. The criterion proposes, first, to write $P$ in its disjunctive normal form (DNF) \cite{Fitting96}:
\begin{center}
$P=(P_1^1 \wedge \ldots \wedge P_{n_1}^1) \vee (P_1^2 \wedge \ldots \wedge P_{n_2}^2) \vee \ldots \vee (P_1^m \wedge \ldots \wedge P_{n_m}^m)$
\end{center}
and, afterward, partition the simulation configurations set according to this DNF:
\begin{center}
$\begin{aligned}
IniSt_i &= \{s \in S \mid (P_1^i(s) \wedge \ldots \wedge P_{n_i}^i(s))\}, \\
InPairs_i &= InPairs.
\end{aligned}$
\end{center}
with $i \in [1, m]$. The same idea can be applied to the set of inputs, outputs or any other intentional set of the model.

For instance, let us suppose that the set of inputs $X$ of a given model is $X=\{(n,m) | n*m>0 \Rightarrow n>m\}$. Therefore, the predicate $P=n*m>0 \Rightarrow n>m$ is written is its DNF by applying a known algorithm \cite{Fitting96}, $P=\neg(n*m>0) \vee (n>m)$ and two SCCs should be defined, one for the predicate $\neg(n*m>0)$ and the other for $n>m$.

\subsubsection{Standard Partitions}
\label{StanPart}
In almost all models, different mathematical operators appear in the definitions of the model elements (transition functions, time advance function, state values) and they can be simple (addition, sets union) or more complex (operators defined in terms of simpler ones, functions defined in a programming language or in a pseudo-code). Each operator has a particular input domain and this criterion proposes to divide this domain associating to it a \textit{standard partition}. A standard partition is a partition of the operator's domain into sets called sub-domains; each sub-domain is defined by the conditions that each operand of the operation must satisfy. Thus, each sub-domain is transformed into a condition to generate a SCC.

Therefore, for each operator in the model a standard partition should be defined. For example, for the operator $<$ ($a$ $<$ $b$), the standard partition could be \cite{CM09}:

\begin{center}
\begin{tabular}{c c c}
$a<0, b<0$ & $a<0, b=0$ & $a<0, b>0$ \\
$a=0, b<0$ & $a=0, b=0$ & $a=0, b>0$ \\
$a>0, b<0$ & $a>0, b=0$ & $a>0, b>0$ \\
\end{tabular}
\end{center}

Let us suppose that $x_1, \ldots, x_n$ are some of the variables that define the set of states, $S$, of some model and $\theta(x_1, \ldots, x_n)$ is an operator of arity $n$ with the associated standard partition $SP_1(x_1, \ldots , x_n)$, $\ldots,$ $SP_m(x_1, \ldots, x_n)$. When $\theta$ appears in an expression the set of possible simulations must be partitioned using the standard partition associated with it:
\begin{center}
$\begin{aligned}
IniSt_i &= \{s \in S \mid SP_i(x_1, \ldots, x_n)\},\\
InPairs_i &= InPairs.
\end{aligned}$
\end{center}

\subsubsection{Domain Propagation}
This is a particular criterion, since it does not generate new partitions by itself. The purpose of it is to obtain standard partitions of complex operators combining the standard partitions of simpler sub-operators. 

Each sub-operation has input domain partitions of its own which are ignored by the standard partitions criterion if it is applied to the complex operator. Using domain propagation the input domain partition of sub-operations are propagated to the higher level \cite{Stocks93}.

For example, let $\square$ be a complex operator defined as: $\square(A,B,C) = (A\triangle B)\Diamond C$ where $\triangle$ and $\Diamond$ are simple operators.

Let us suppose that $\triangle$ and $\Diamond$ have the following standard partitions:

\begin{center}
$SP^\triangle(S,T)=D_1^\triangle(S,T) \vee \ldots \vee D_n^\triangle(S,T)$ \\
$SP^\Diamond(U,V)=D_1^\Diamond(U,V) \vee \ldots \vee D_k^\Diamond(U,V)$
\end{center}

We apply first $SP^\triangle$ to the sub-expression $(A\triangle B)$, replacing the formal parameters appearing in $SP^\triangle$ by $A$ and $B$ respectively:
\begin{center}
$SP^\triangle(A,B)=D_1^\triangle(A,B) \vee \ldots \vee D_m^\triangle(A,B)$ \\
\end{center}
with $m \leq n$.

Afterward we do the same with $SP^\Diamond$, obtaining:
\begin{center}
$SP^\Diamond(A\triangle B, C)=D_1^\Diamond(A\triangle B, C), \vee \ldots \vee D_j^\Diamond(A\triangle B, C)$
\end{center}
with $j \leq k$.

Finally, we combine both propositions obtained and simplify:
\begin{center}
$SP^\square = SP^\triangle(A,B) \wedge SP^\Diamond(A\triangle B, C)$
\end{center}

\subsubsection{Time Partitions}
In timed formalisms, like DEVS, modeling the time is a crucial issue. It is very common, in DEVS models, to use additional variables to model time. Furthermore, one characteristic of these models is that the elapsed time appears in the external transition function definition as a variable. Therefore, in the validation of such a model, a question arise: \textit{``how do we know if each event has been already simulated in all relevant or significant time interval?''}. That is, those time intervals in which it is possible to find an error in the model. Again, to answer that question all events must be simulated in all possible time interval making the validation an infinite process.

Therefore, this criterion consists in, first, identifying those variables of the state that interact with the elapsed time or are used to simulate the time advance or timers, for instance. Afterward it defines key time points, or intervals according with the interaction of those variables, for example, using the standard partition for the operations that involve every time variable. Once these key time intervals are defined, one class must be generated for each input at each time point. For example, let us suppose that the time interval $[a,b]$ is relevant to see a particular functionality of the model, therefore it would be meaningful to simulate input events $(x,t)$ with $x \in X \cup \{\tau\}$ and times $t<a$, $t=a$, $a<t<b$, $t=b$ and $t>b$. Formally, for each defined time interval $[a_i, b_i]$, five SCCs should be created:
\begin{itemize}
    \item $IniSt_i^1 = \{s \in S\}$,\\
          $InPairs_i^1 = \{(x,t) \mid x \in X \cup \{\tau\}, t \in \mathbb{R}^+_0 \wedge t < a_i\}$.
    \item $IniSt_i^2 = \{s \in S\}$,\\
          $InPairs_i^2 = \{(x,a_i) \mid x \in X \cup \{\tau\}\}$.
    \item $IniSt_i^3 = \{s \in S\}$,\\
          $InPairs_i^3 = \{(x,t) \mid x \in X \cup \{\tau\}, t \in \mathbb{R}^+_0 \wedge a_i < t < b_i\}$.
    \item $IniSt_i^4 = \{s \in S\}$,\\
          $InPairs_i^4 = \{(x,b_i) \mid x \in X \cup \{\tau\}\}$.
    \item $IniSt_i^5 = \{s \in S\}$,\\
          $InPairs_i^5 = \{(x,t) \mid x \in X \cup \{\tau\}, t \in \mathbb{R}^+_0 \wedge t > b_i\}$.
\end{itemize}
and for each defined time point $t_j$, three classes should be created:
\begin{itemize}
    \item $IniSt_j^1 = \{s \in S\}$,\\
          $InPairs_j^1 = \{(x,t) \mid x \in X \cup \{\tau\}, t \in \mathbb{R}^+_0 \wedge t < t_j\}$.
    \item $IniSt_j^2 = \{s \in S\}$,\\
          $InPairs_j^2 = \{(x,t_j) \mid x \in X \cup \{\tau\}\}$.
    \item $IniSt_j^3 = \{s \in S\}$,\\
          $InPairs_j^3 = \{(x,t) \mid x \in X \cup \{\tau\}, t \in \mathbb{R}^+_0 \wedge t > t_j\}$.
\end{itemize}

The intention of this criterion is to simulate different scenarios where events occur at different moments, to verify the interaction of those variables used to simulate the time among them and the interaction between these ones with the elapsed time (for external transitions).

Later, when this classes are combined with those generated, for instance, by the Extensional Sets criterion, it will be able to simulate all input events in all relevant time instants.

\subsection{Combining Classes}
\label{CombPart}
As has been mentioned in Section \ref{PartCrit}, the partition criteria are independent from each other. Furthermore, each simulation configuration aims to validate a particular characteristic of the model. However, it is usually more efficient to simultaneously validate more than one characteristic. We use efficiency as a measure of the number of errors found in the model. Computational efficiency is not addressed here.

In order to achieve this, it is better to combine the classes generated by applying each criterion. Observe that, however, not all criteria are always applied. This will depend on the model and on the time available for validating it. Also, observe that the same criterion can be applied more than once on the same model. For instance, if the criterion known as Extensional Sets is firstly applied over the state set and secondly over the input set of a given model, the result is two independent sets of configurations. These sets of configurations can be combined in order to simulate the arrival of every input event on every state. In this way, these combined SCCs would find errors, for example, due to the arrival of an event on a wrong state.

Let us see how SCCs can be combined with a toy example. Let $M_T = (X, Y, S, \delta_{int}, \delta_{ext}, \lambda, ta)$, with $X=\mathbb{N}$ and $S=\mathbb{N} \times \{ON, OFF\}$, be part of the model of some system. Suppose that after applying some criteria the following SCCs are obtained:
\begin{itemize}
    \item $SCC_a$: \\
		  $IniSt_a$ = $\{(n,m) \in S \mid n \leq 10 \}$,\\
          $InPairs_a$ = $\{(1, t) \mid t \in \mathbb{R}_0^+\}$
    \item $SCC_b$: \\
		  $IniSt_b$ = $\{(n,m) \in S \mid m = ON \}$,\\
          $InPairs_b$ = $\{(1, t) \mid t \in \mathbb{R}_0^+\}$
    \item $SCC_c$: \\
		  $IniSt_c$ = $\{(n,m) \in S \mid m = OFF \}$,\\
          $InPairs_c$ = $\{(1, t) \mid t \in \mathbb{R}_0^+\}$
\end{itemize}
Now, we can combine these classes by intersecting the corresponding $IniSt$ and $InPairs$ sets as follows:
\begin{itemize}
    \item $SCC_d = SCC_a \wedge SCC_b$:
    \vspace{-0.2cm}
    \begin{align*}
		IniSt_d &= IniSt_a \cap IniSt_b = \{(n,m) \in S \mid n \leq 10 \} \cap \{(n,m) \in S \mid m = ON \} = \\
			&= \{(n,m) \in S \mid n \leq 10 \wedge m = ON \},
	\end{align*}
	\vspace{-0.9cm}
	\begin{align*}
        InPairs_d &= InPairs_a \cap InPairs_b = \{(1, t) \mid t \in \mathbb{R}_0^+\} \cap \{(1, t) \mid t \in \mathbb{R}_0^+\} = \\
			&= \{(1, t) \mid t \in \mathbb{R}_0^+\}
	\end{align*}
	\item $SCC_e = SCC_a \wedge SCC_c$:
	\vspace{-0.2cm}
	\begin{align*}
          IniSt_e &= IniSt_a \cap IniSt_c = \{(n,m) \in S \mid n \leq 10 \} \cap \{(n,m) \in S \mid m = OFF \} = \\ 
          &= \{(n,m) \in S \mid n \leq 10 \wedge  m = OFF \},
	\end{align*}
	\vspace{-0.9cm}
	\begin{align*}
          InPairs_e &= InPairs_a \cap InPairs_c = \{(1, t) \mid t \in \mathbb{R}_0^+\} \cap \{(1, t) \mid t \in \mathbb{R}_0^+\} = \\
          &= \{(1, t) \mid t \in \mathbb{R}_0^+\}
	\end{align*}
	\item $SCC_f = SCC_b \wedge SCC_c$:
	\vspace{-0.2cm}
	\begin{align*}
		  IniSt_f &= IniSt_b \cap IniSt_c = \{(n,m) \in S \mid m = ON \} \cap \{(n,m) \in S \mid m = OFF \} = \\
		  &= \{\}
	\end{align*}
	\vspace{-0.9cm}
	\begin{align*}
		  InPairs_f &= InPairs_b \cap InPairs_c = \{(1, t) \mid t \in \mathbb{R}_0^+\} \cap \{(1, t) \mid t \in \mathbb{R}_0^+\} = \\
          &= \{(1, t) \mid t \in \mathbb{R}_0^+\}
	\end{align*}
\end{itemize}
Note, however, that only $SCC_d$ and $SCC_e$ are valid SCC since $SCC_f$ is empty because $IniSt_f$ is an empty set.

As this example shows, combining SCCs by intersection may yield empty classes. If this is the case: a) eliminate them; and b) keep the SCCs that produced the empty ones.

Observe that, since intersection is commutative, it is the same to combine, for instance, $SCC_a$ and $SCC_b$ as $SCC_a \cap SCC_b$ or as $SCC_b \cap SCC_a$. Moreover, if $SCC_d$ is the result of combining $SCC_a$ and $SCC_b$, and $SCC_d$ is combined with some $SCC_{\alpha}$ the result is $SCC_b \cap SCC_a \cap SCC_{\alpha}$ which is the same regardless of the order in which classes are combined. In summary, SCCs can be combined in any order.

\subsection{Simulation Sequencing}
Once all SCCs are defined it is necessary to set the initial state of the simulation and to execute a transition. However, the new state obtained from the transition may be the initial state of another SCC. If this is the case, then it would be computationally more efficient to continue the simulation by executing the transition indicated by this other SCC. This avoids a new configuration of another simulation for the mentioned SCC, i.e. this reuses the configuration left by the previous run. Then, this process yields a sequence of configurations that should be run one after the other.

Here we propose an algorithm to generate these sequences by analyzing the classes obtained by the application of the criteria. The pseudo code of the algorithm can be seen in Algorithm \ref{Alg:Sec}, and it is explained below.

\begin{algorithm}[h]
\caption{Simulation Sequencing}
\label{Alg:Sec}
\begin{algorithmic}[1]
\REQUIRE{$TotSCC$: Set of all SCCs generated by applying our methods}
\ENSURE{$SimSeq$: Set of configuration simulation sequences}
\WHILE{$TotSCC \neq \emptyset$}
    \STATE select $scc \in TotSCC$
    \STATE select $is \in scc.IniSt$
    \STATE select $ip \in scc.InPair$
    \STATE $s \leftarrow simulate(is,ip)$
    \STATE $seq \leftarrow (is,ip)$
    \STATE $TotSCC \leftarrow TotSCC \setminus \{scc\}$
    \WHILE{$\exists \ scc' \in TotSCC \mid s \in scc'.IniSt$}
        \STATE select $scc' \in TotSCC \mid s \in scc'.IniSt$
        \STATE select $ip' \in scc'.InPairs$
        \STATE $s \leftarrow simulate(s,ip')$
        \STATE $seq \leftarrow seq \circ (s,ip')$
        \STATE $TotSCC \leftarrow TotSCC \setminus \{scc'\}$
    \ENDWHILE
    \STATE $SimSeq \leftarrow SimSeq \cup \{seq\}$
\ENDWHILE
\end{algorithmic}
\end{algorithm}

The main progress of Algorithm \ref{Alg:Sec} is led by the following idea: select one SCC ($scc \in TotSCC$), an initial state from that SCC ($is \in scc.IniSt$) and simulate one event of the set of input pairs $(event, time)$ associated to that SCC ($ip \in scc.InPair$).  $simulate(s,ip')$ means to execute on step of the simulation model. Once the selected event has been simulated and a new state, $s$, is reached the current sequence is updated (sentence 6) and the set of SCCs is reduced by removing the SCC added to the sequence (sentence 7). Now, there are two alternatives: a) wait for an internal transition to occur (if $ta(s) \neq \infty$); b) simulate another event. As shown in sentence 8, if there is a SCC, $scc'$, such that $s$ is one of its initial states, then $scc'$ is chosen as the next SCC. In which case one of its input pairs is simulated (sentences 10 and 11). Afterward, the process continues repeatedly choosing an event or waiting for an internal transition. At any point of this process, if the state reached from the last simulation step does not belong to any initial state of the remaining SCCs, that sequence ends there and it is added to the set of configuration simulation sequences (sentence 15). A new configuration simulation sequence is started if there are more SCCs.

Thereby, with Algorithm \ref{Alg:Sec} all SCCs are used at least once. Further, more complex coverage criteria could be defined for the generation of the sequences, for example, in a similar way than Souza et al. \cite{SMFM00}.

\section{Discussion and Automation}
\label{Auto}
As has been mentioned in the introduction, the technique presented in this paper would allow the automation of an important part of the DEVS model validation process. The intention of this section is to show that it would be possible to build a software tool to assist engineers when they apply our method. Therefore, we explain what can be automatically done in each step and we describe the main issues that need to be addressed in order to achieve this automation. Also, further considerations about this methodology are discussed.

An overview of this semi-automated process is the following:
\begin{enumerate}
 \item Parse the mathematical description of the DEVS model.
 \item Select and apply the partition criteria to generate the simulations.
 \item Select simulations and generate simulation sequences.
 \item Translate the simulation configurations into some simulation language and simulate them.
 \item Translate the simulation results into the model formalism.
 \item Compare the results with the requirements in order to achieve a verdict about the validation of the model.
\end{enumerate}

The mathematical description of a DEVS model can be automatically be parsed only if it is written in a standard, formal notation. Being so important this issue in the modeling and simulation community, there exists an international group \cite{StDEVS} trying to develop standards for a computer processable representation of the mathematical description of DEVS models. This is still an open area. This standard should be based on mathematics and logics, for instance, like Z \cite{Spivey89} or TLA \cite{Lamport94}, instead of a standard based on or akin to a programming language. In this way, engineers do not need to have programming skills for writing their models. 
For example, we think that DEVSpecL \cite{HK06}, a DEVS specification language developed by Hong and Kim, looks more like a programming language rather than a mathematical representation of DEVS. The definition of a standard mathematical language for DEVS involves the definition of its syntax and semantic and the implementation of a compiler or parser to transform the mathematical model to the input language of a simulator.

Regarding the application of the criteria, a preliminary analysis indicates that it would be appropriate to apply them semi-automatically. We think that a tool should allow engineers to select which criterion must be applied to which part of the model. Moreover, the engineer could add new criteria and use them. Another alternative would be to implement an heuristic that automatically select the criteria according to an analysis over the description of the model. 

The application of the criteria involves an important problem, which is the total number of SCCs generated. A preliminary analysis indicates that, although the number of classes could be considerable, it is not exponential since the number of criteria involved is fixed and small. The crucial issue is the combination between classes. The order of classes is given by $O(x^n)$ where $x$ is the number of classes generated by a criterion and $n$ the number of criteria applied (recall that $n$ is fixed and small).

Once the criteria have been applied and the SCCs have been defined in the following phase one simulation configuration for each SCC is selected. Finding a simulation configuration for a given SCC means to find an element belonging to it. Usually, this involves solving a formula over Set Theory, Arithmetic over Integer and Real Numbers and, possibly, other mathematical theories. Further, free variables range over infinite sets, making the problem undecidable. One possibility is to adopt a Satisfiability Modulo Theories (SMT) solver \cite{GKSS08,NOT06}, by adapting the work of Cristi\'{a} and Frydman \cite{CF12}. Another alternative could be to use constraint solvers such as $\{log\}$ (pronounced `setlog') \cite{DOPR96,DPPR00,CRF13}.

The simulation configurations generated above, are described essentially in mathematics. Therefore, to perform the simulations these need to be refined, i.e. rewritten in the input language of some simulation tool, as the model definition. A possible way to do this, is adapting the work done for MBT \cite{Cetal11}. This would be a semi-automatic process, since the engineer must define some rules to do this translation.

After performing the simulations, the reverse process must be done. That is, the results of these simulations must be rewritten in the mathematical or formal language used to describe the DEVS model. Again, this would be a semi-automatic process due to the definition of the translation rules. Then, the simulation results can be compared with the expected result (Requirements). This comparison is necessarily manual since involves the requirements.

Besides, it is possible to use this methodology to test the concrete simulation model. However, note that both validations cannot be done simultaneously for the same model. In effect, if our methodology is used to validate the abstract model it is necessary to assume that the concrete model is a faithful representation of the abstract one. On the other hand, if it is used to validate the concrete model it is necessary to assume that the abstract model is correct with respect to the requirements.

Finally, as can be seen, the simulation criteria are applied over the mathematical definition of the model and are based also on mathematical or logical operations. Therefore, the proposed validation methodology is rigorous and systematic (because it is based on mathematics, logic and formal languages). Moreover this methodology is systematic since the whole set  of simulation configurations is systematically partitioned obtaining refined and more expressive simulation configuration classes. In this way, given that covering all possible paths of the model is impossible (because they are infinite) we propose to cover the logical and mathematical structure of the model, thus, covering the significant paths of the model.

\section{Case Studies}
\label{CaseStudies}
In this section we show the application of the criteria in two examples. In each case, first we present the requirements of the system, then we describe the DEVS model and finally the SCCs that follow the application of the criteria.

\subsection{Elevator}
The following requirements correspond to the control system of an elevator. It has a control panel with one button for each floor and two other buttons, one for opening and one for closing the door. Furthermore, it has a switch to interrupt or restart its operation. Every time the elevator reaches a floor it receives a signal (the same signal for all floors). To prevent accidents or malfunction of the elevator, it has two sensors, one to check, before or during the door close, if someone or something is crossing it; and the other one to prevent exceeding the weight limit of the elevator. To complete the functionality of the elevator, the system has three timers, whose purpose is described below.

The requirements of the system are:
\begin{itemize}
    \item The door should not be closed if:
    \begin{itemize}
        \item Any sensor is active (someone or something is crossing the door, or the weight limit is exceeded).
        \item The Stop Switch is on.
    \end{itemize}
    \item If the elevator is stopped at any floor and someone call it from another floor or someone press the button (on the control panel) of another floor, after $T_{D_1}$ units of time the door should start closing, and takes $T_{D_2}$ units of time to close completely. However, if the Open Button is pressed or the door sensor activates the timer is reset.
    \item If after $T_A$ units of time ($T_A > T_{D_1}$) since the elevator is called the door is not closed, an alarm is fired. Unlike the former timer, this is not reset even though the door sensor is activated or the open button is pressed. It is reset only when the door is totally closed and the elevator starts moving. If the alarm is fired, it should be turned off when this timer is reset.
    \item After $T_{GF}$ units of time since the elevator is stopped in a floor different from the ground, if no one calls it, it should return to the ground floor and open its door.
    \item This elevator has no memory, therefore, it goes to the first floor indicated since it has been stopped.
\end{itemize}

\subsubsection{DEVS Model}
Figure \ref{ElevatorDEVS1}, Figure \ref{ElevatorDEVS2} and Figure \ref{ElevatorDEVS3} represent a possible DEVS model corresponding to the control system of the elevator described before.

\begin{figure}[H]
\hrulefill

\sffamily
\begin{scriptsize}
$M_{sv}$ = $(S,X,Y,\delta_{int}, \delta_{ext}, \lambda, ta)$

\vspace{0.1cm}
$S=ActualFloor \times FloorCalled \times Engine \times Door \times Sensors \times Switch \times Alarm \times Timers \times NextTimer$

\hspace{.6cm}where:

\hspace{.6cm}$ActualFloor = \mathbb{N}$

\hspace{.6cm}$FloorCalled = \mathbb{N} \cup \{\emptyset\}$

\hspace{.6cm}$Engine = \{\mathsf{up}$, $\mathsf{down}$, $\mathsf{stopped}\}$

\hspace{.6cm}$Door = \{\mathsf{open}$, $\mathsf{closed}$, $\mathsf{closing}\}$

\hspace{.6cm}$Sensors = \{0,1\} \times \{0,1\}$

\hspace{.6cm}$Alarm = Switch = \{0,1\}$

\hspace{.6cm}$Timers = (\mathbb{R}_0^+ \cup \{\infty\}) \times (\mathbb{R}_0^+ \cup \{\infty\}) \times (\mathbb{R}_0^+ \cup \{\infty\}) \times (\mathbb{R}_0^+ \cup \{\infty\}) \times (\mathbb{R}_0^+ \cup \{\infty\})$

\hspace{.6cm}$NextTimer = \{\mathsf{A}$, $\mathsf{D_1}$, $\mathsf{D_2}$, $\mathsf{GF}$, $\mathsf{O}\}$

\vspace{0.1cm}
$X$ = $\mathbb{N}$ $\cup$ $\{\mathsf{fsig}$, $\mathsf{ws_{on}}$, $\mathsf{ws_{off}}$, $\mathsf{ds_{on}}$, $\mathsf{ds_{off}}$, $\mathsf{od_{press}}$, $\mathsf{cd_{press}}$, $\mathsf{s_{on}}$, $\mathsf{s_{off}}\}$

\vspace{0.1cm}
$Y=(\mathbb{N} \cup \{\mathsf{ST}\}) \times \{\mathsf{up}$, $\mathsf{down}$, $\mathsf{stop}$, $\bot\}$ $\times$ $\{\mathsf{opendoor}$, $\mathsf{closedoor}$, $\bot\}$ $\times$ $\{\mathsf{firealarm}$, $\mathsf{stopalarm}$, $\bot\}$

\vspace{0.1cm}
$\delta_{int}((f, fc, eng, d, (ws, ds), sw, a, (at, dt1, dt2, gft, ot),nt)) = $
\vspace{-0.3cm}
\begin{numcases}{}
	(f, \emptyset, \mathsf{stopped}, \mathsf{open}, (ws, ds), sw, a, (\infty, \infty, \infty, \mathsf{T_{GF}}, \infty), nt'(\infty, \infty, \infty, \mathsf{T_{GF}}, \infty)) \nonumber\\
    \hspace{0.5cm} \text{if } nt = \mathsf{O} \wedge eng \neq \mathsf{stopped} \wedge f = fc \wedge f \neq 0\\
    (f, \emptyset, \mathsf{stopped}, \mathsf{open}, (ws, ds), sw, a, (\infty, \infty, \infty, \infty, \infty), nt'(\infty, \infty, \infty, \mathsf{T_{GF}}, \infty)) \nonumber\\
    \hspace{0.5cm} \text{if } nt = \mathsf{O} \wedge eng \neq \mathsf{stopped} \wedge f = fc \wedge f=0\\
    (f, fc, eng, d, (ws, ds), sw, a, (\infty, \infty, \infty, \infty, \infty), nt'(\infty, \infty, \infty, \infty, \infty)) \nonumber\\
    \hspace{0.5cm} \text{if } nt = \mathsf{O} \wedge eng \neq \mathsf{stopped} \wedge f \neq fc \\
    (f, fc, \mathsf{stopped}, d, (ws, ds), sw, a, (\mathsf{T_A}, \infty, \infty, \infty, \infty), nt'(\mathsf{T_A}, \infty, \infty, \infty, \infty)) \nonumber\\
    \hspace{0.5cm} \text{if } nt=\mathsf{O} \wedge sw=1 \wedge eng \neq \mathsf{stopped}\\
    (f, fc, eng, d, (ws, ds), sw, a, (at-ot, dt1-ot, dt2-ot, gft-ot, \infty), nt'(at-ot, dt1-ot, dt2-ot, gft-ot, \infty)) \nonumber\\
    \hspace{0.5cm} \text{if } nt=\mathsf{O} \wedge sw=1 \wedge eng = \mathsf{stopped}\\
    (f, fc, \mathsf{up}, d, (ws, ds), sw, 0, (\infty, \infty, \infty, \infty, \infty), nt'(\infty, \infty, \infty, \infty, \infty)) \nonumber\\
    \hspace{0.5cm} \text{if } nt=\mathsf{O} \wedge ds=0 \wedge ws=0 \wedge sw=0 \wedge d=\mathsf{closed} \wedge fc \neq \emptyset \wedge fc > f \\
    (f, fc, \mathsf{down}, d, (ws, ds), sw, 0, (\infty, \infty, \infty, \infty, \infty), nt'(\infty, \infty, \infty, \infty, \infty)) \nonumber\\
    \hspace{0.5cm} \text{if } nt=\mathsf{O} \wedge ds=0 \wedge ws=0 \wedge sw=0 \wedge d=\mathsf{closed} \wedge fc \neq \emptyset \wedge fc < f \\
    (f, fc, eng, \mathsf{closing}, (ws, ds), sw, 0, (at-ot, \infty, \mathsf{T_{D_2}}, \infty, \infty), nt'(at-ot, \infty, \mathsf{T_{D_2}}, \infty, \infty)) \nonumber\\
    \hspace{0.5cm} \text{if } nt=\mathsf{O} \wedge ds=0 \wedge ws=0 \wedge sw=0 \wedge d=\mathsf{open} \wedge fc \neq \emptyset \\
    (f, fc, eng, d, (ws, ds), sw, a, (\infty, \infty, \infty, gft-ot, \infty), nt'(\infty, \infty, \infty, gft-ot, \infty)) \nonumber\\
    \hspace{0.5cm} \text{if } nt=\mathsf{O} \wedge ds=0 \wedge ws=0 \wedge sw=0 \wedge fc = \emptyset \\
    (f, fc, eng, \mathsf{open}, (ws, ds), sw, a, (at-ot, \mathsf{T_{D_1}}, \infty, \infty, \infty), nt'(at-ot, \mathsf{T_{D_1}}, \infty, \infty, \infty)) \nonumber\\
    \hspace{0.5cm} \text{if } nt=\mathsf{O} \wedge d=\mathsf{closing} \\
    (f, fc, eng, \mathsf{closing}, (ws, ds), sw, a, (at-dt1, \infty, \mathsf{T_{D_2}}, \infty, ot-dt1), nt'(at-dt1, \infty, \mathsf{T_{D_2}}, \infty, ot-dt1)) \nonumber\\
    \hspace{0.5cm} \text{if } nt=\mathsf{D_1} \wedge ds=0 \wedge ws=0 \wedge sw=0 \\
    (f, fc, eng, d, (ws, ds), sw, a, (at-dt1, \infty, \infty, \infty, ot-dt1), nt'(at-dt1, \infty, \infty, \infty, ot-dt1)) \nonumber\\
    \hspace{0.5cm} \text{if } nt=\mathsf{D_1} \neg (\wedge ds=0 \wedge ws=0 \wedge sw=0) \\ 
    (f, fc, \mathsf{up}, \mathsf{closed}, (ws, ds), sw, 0, (\infty, \infty, \infty, \infty, ot-dt2), nt'(\infty, \infty, \infty, \infty, ot-dt2)) \nonumber\\
    \hspace{0.5cm} \text{if } nt=\mathsf{D_2} \wedge ds=0 \wedge ws=0 \wedge sw=0 \wedge fc > f \\
    (f, fc, \mathsf{down}, \mathsf{closed}, (ws, ds), sw, 0, (\infty, \infty, \infty, \infty, ot-dt2), nt'(\infty, \infty, \infty, \infty, ot-dt2)) \nonumber\\
    \hspace{0.5cm} \text{if } nt=\mathsf{D_2} \wedge ds=0 \wedge ws=0 \wedge sw=0 \wedge fc < f \\
    (f, fc, eng, d, (ws, ds), sw, a, (at-dt2, \infty, \infty, \infty, ot-dt2), nt'(at-dt2, \infty, \infty, \infty, ot-dt2)) \nonumber\\
    \hspace{0.5cm} \text{if } nt=\mathsf{D_2} \wedge \neg (ds=0 \wedge ws=0 \wedge sw=0) \\ 
    (f, fc, eng, d, (ws, ds), sw, 1, (\infty, dt1-at, dt2-at, gft-at, ot-at), nt'(\infty, dt1-at, dt2-at, gft-at, ot-at)) \nonumber\\
    \hspace{0.5cm} \text{if } nt=\mathsf{A} \\
    (f, 0, eng, \mathsf{closing}, (ws, ds), sw, a, (\infty, \infty, \mathsf{T_{D_2}}, \infty, ot-gft), nt'(\infty, \infty, \mathsf{T_{D_2}}, \infty, ot-gft)) \nonumber\\
    \hspace{0.5cm} \text{if } nt=\mathsf{GF} \wedge f \neq 0 \wedge fc = \emptyset \wedge d=\mathsf{open} \wedge ds=0 \wedge ws=0 \wedge sw=0 \\
    (f, 0, eng, d, (ws, ds), sw, a, (at-gft, \infty, \infty, \infty, ot-gft), nt'(at-gft, \infty, \infty, \infty, ot-gft)) \nonumber\\
    \hspace{0.5cm} \text{if } nt=\mathsf{GF} \wedge f \neq 0 \wedge fc = \emptyset \wedge d=\mathsf{open} \wedge \neg (ds=0 \wedge ws=0 \wedge sw=0)
\end{numcases}
\vspace{-0.3cm}
\dotfill
\end{scriptsize}
\caption{DEVS Model of an elevator control system (part a)}
\label{ElevatorDEVS1}
\end{figure}

\begin{figure}[H]
\sffamily
\begin{scriptsize}
\dotfill

$\delta_{ext}((f, fc, eng, d, (ws, ds), sw, a, (at, dt1, dt2, gft, ot), nt), e, x) = $
\setcounter{equation}{0}
\begin{numcases}{}
    (f, n, eng, d, (ws, ds), sw, a, (\mathsf{T_A}, \mathsf{T_{D_1}}, dt2', \infty, ot'), nt'(\mathsf{T_A}, \mathsf{T_{D_1}}, dt2', \infty, ot')) \nonumber\\
    \hspace{1cm} \text{if } x=n, n \in \mathbb{N} \wedge n \neq f \wedge eng = \mathsf{stopped} \wedge fc = \emptyset \\
    (f+1, fc, eng, d, (ws, ds), sw, a, (at', dt1', dt2', gft', 0), nt'(at', dt1', dt2', gft', 0)) \nonumber\\
    \hspace{1cm} \text{if } x=\mathsf{fsig} \wedge eng=\mathsf{up} \\
    (f-1, fc, eng, d, (ws, ds), sw, a, (at', dt1', dt2', gft', 0), nt'(at', dt1', dt2', gft', 0)) \nonumber\\
    \hspace{1cm} \text{if } x=\mathsf{fsig} \wedge eng=\mathsf{down}\\
    (f, fc, eng, d, (ws, 1), sw, a, (at', dt1', dt2', gft', 0), nt'(at', dt1', dt2', gft', 0)) \nonumber\\
    \hspace{1cm} \text{if } x=\mathsf{ds_{on}} \wedge eng=\mathsf{stopped}))\\
    (f, fc, eng, d, (ws, 1), sw, a, (at', dt1', dt2', gft', ot'), nt'(at', dt1', dt2', gft', ot')) \nonumber\\
    \hspace{1cm} \text{if } x=\mathsf{ds_{on}} \wedge eng \neq \mathsf{stopped}))\\
    (f, fc, eng, d, (ws, 0), sw, a, (at', \mathsf{T_{D_1}}, dt2', gft', ot'), nt'(at', \mathsf{T_{D_1}}, dt2', gft', ot')) \nonumber\\
    \hspace{1cm} \text{if } x=\mathsf{ds_{off}} \wedge d=\mathsf{open} \wedge fc \neq \emptyset \wedge ws=0 \wedge sw=0\\
    (f, fc, eng, d, (ws, 0), sw, a, (at', dt1', dt2', gft', ot'), nt'(at', dt1', dt2', gft', ot')) \nonumber\\
    \hspace{1cm} \text{if } x=\mathsf{ds_{off}} \wedge (d \neq \mathsf{open} \vee fc=\emptyset \vee ws=1 \vee sw=1)\\
    (f, fc, eng, d, (1, ds), sw, a, (at', dt1', dt2', gft', 0), nt'(at', dt1', dt2', gft', 0)) \nonumber\\
    \hspace{1cm} \text{if } x=\mathsf{ws_{on}} \wedge eng=\mathsf{stopped}\\
    (f, fc, eng, d, (1, ds), sw, a, (at', dt1', dt2', gft', ot'), nt'(at', dt1', dt2', gft', ot')) \nonumber\\
    \hspace{1cm} \text{if } x=\mathsf{ws_{on}} \wedge eng \neq \mathsf{stopped}\\
    (f, fc, eng, d, (0, ds), sw, a, (at', \mathsf{T_{D_1}}, gft', ot'), nt'(at', \mathsf{T_{D_1}}, gft', ot')) \nonumber\\
    \hspace{1cm} \text{if } x=\mathsf{ws_{off}} \wedge fc \neq \emptyset \wedge d=\mathsf{open} \\
    (f, fc, eng, d, (0, ds), sw, a, (at', dt1', dt2', gft', ot'), nt'(at', dt1', dt2', gft', ot')) \nonumber\\
    \hspace{1cm} \text{if } x=\mathsf{ws_{off}} \wedge (fc = \emptyset \vee d \neq \mathsf{open}) \\
    (f, fc, eng, d, (ws, ds), 1, a, (at', dt1', dt2', gft', 0), nt'(at', dt1', dt2', gft', 0)) \nonumber\\
    \hspace{1cm} \text{if } x=\mathsf{s_{on}} \\
    (f, fc, eng, d, (ws, ds), 0, a, (at', \mathsf{T_{D_1}}, dt2', gft', ot'), nt'(at', \mathsf{T_{D_1}}, dt2', gft', ot')) \nonumber\\
    \hspace{1cm} \text{if } x=\mathsf{s_{off}} \wedge d = \mathsf{open} \wedge fc \neq \emptyset \wedge fc \neq f \\
    (f, fc, eng, d, (ws, ds), 0, a, (at', dt1', dt2', gft', ot'), nt'(at', dt1', dt2', gft', ot')) \nonumber\\
    \hspace{1cm} \text{if } x=\mathsf{s_{off}} \wedge fc = \emptyset \\
    (f, fc, eng, d, (ws, ds), 0, a, (at', dt1', dt2', gft', 0), nt'(at', dt1', dt2', gft', 0)) \nonumber\\
    \hspace{1cm} \text{if } x=\mathsf{s_{off}} \wedge fc \neq \emptyset \wedge d=\mathsf{closed} \\
    (f, fc, eng, d, (ws, ds), sw, a, (at', dt1', dt2', gft', 0), nt'(at', dt1', dt2', gft', 0)) \nonumber\\
    \hspace{1cm} \text{if } x=\mathsf{od_{press}} \wedge d=\mathsf{closing} \\
    (f, fc, eng, d, (ws, ds), sw, a, (at', 0, dt2', gft', ot'), nt'(at', 0, dt2', gft', ot')) \nonumber\\
    \hspace{1cm} \text{if } x=\mathsf{cd_{press}} \wedge d=\mathsf{open} \wedge fc \neq \emptyset \wedge ds = 0 \wedge ws = 0 \wedge sw = 0 \\
    (f, fc, eng, d, (ws, ds), sw, a, (at', dt1', dt2', gft', ot'), nt'(at', dt1', dt2', gft', ot')) \nonumber\\
    \hspace{1cm} \text{Otherwise} 
\end{numcases}

\vspace{0.1cm}
$nt'(at,dt1,dt2,gft,ot)=$
$\begin{cases}
    \mathsf{A}   & \text{if } min(at,dt1,dt2,gft,ot) = at \\
    \mathsf{D_1} & \text{if } min(at,dt1,dt2,gft,ot) = dt1 \\
    \mathsf{D_2} & \text{if } min(at,dt1,dt2,gft,ot) = dt2 \\
    \mathsf{GF}  & \text{if } min(at,dt1,dt2,gft,ot) = gft \\
    \mathsf{O}   & \text{if } min(at,dt1,dt2,gft,ot) = ot \\
\end{cases}$

\vspace{0.1cm}
$at'=at-e$, $dt1'=dt1-e$, $dt2'=dt2-e$, $gft'=gft-e$, $ot'=ot-e$,

\vspace{-0.3cm}
\dotfill
\end{scriptsize}
\caption{DEVS Model of an elevator control system (part b)}
\label{ElevatorDEVS2}
\end{figure}

\begin{figure}[H]
\sffamily
\begin{scriptsize}
\dotfill

$\lambda((f, fc, eng, d, (ws, ds), sw, a, (at, dt1, dt2, gft, ot), nt)) = $
\setcounter{equation}{0}
\begin{numcases}{}
    (f, \bot, \mathsf{closedoor}, \bot) & if $\ nt=\mathsf{D_1} \wedge ds=0 \wedge ws=0 \wedge sw=0 $ \\
    (f, \bot, \bot, \bot) & if $\ nt=\mathsf{D_1} \wedge (ws=1 \vee ds=1) \wedge sw=0$ \\
    (ST, \bot, \bot, \bot) & if $\ nt=\mathsf{D_1} \wedge sw=1$ \\
    (f, \mathsf{up}, \bot, \bot) & if $\ nt=\mathsf{D_2} \wedge ds=0 \wedge ws=0 \wedge sw=0 \wedge fc > f \wedge a=0$ \\
    (f, \mathsf{down}, \bot, \bot) & if $\ nt=\mathsf{D_2} \wedge ds=0 \wedge ws=0 \wedge sw=0 \wedge fc < f \wedge a=0$ \\
    (f, \mathsf{up}, \bot, \mathsf{stopalarm}) & if $\ nt=\mathsf{D_2} \wedge ds=0 \wedge ws=0 \wedge sw=0 \wedge fc > f \wedge a=1$ \\
    (f, \mathsf{down}, \bot, \mathsf{stopalarm}) & if $\ nt=\mathsf{D_2} \wedge ds=0 \wedge ws=0 \wedge sw=0 \wedge fc < f \wedge a=1$ \\
    (f, \bot, \bot, \bot) & if $\ nt=\mathsf{D_2} \wedge (ws=1 \vee ds=1) \wedge sw=0$ \\
    (ST, \bot, \bot, \bot) & if $\ nt=\mathsf{D_2} \wedge sw=1$ \\
    (f, \bot, \mathsf{closedoor}, \bot) & if $\ nt=\mathsf{GF} \wedge f \neq 0 \wedge fc=\bot \wedge d=\mathsf{open} \wedge ds=0 \wedge ws=0 \wedge sw=0$ \\
    (f, \bot, \bot, \bot) & if $\ nt=\mathsf{GF} \wedge (f=0 \vee ws=1 \vee ds=1) \wedge sw=0$ \\
    (ST, \bot, \bot, \bot) & if $\ nt=\mathsf{GF} \wedge sw=1$ \\
    (f, \mathsf{stop}, \mathsf{opendoor}, \bot) & if $\ nt=\mathsf{O} \wedge eng \neq \mathsf{stopped} \wedge f = fc$ \\
    (f, \bot, \bot, \bot) & if $\ nt=\mathsf{O} \wedge eng \neq \mathsf{stopped} \wedge f \neq fc$ \\
    (f, \bot, \mathsf{opendoor}, \bot) & if $\ nt=\mathsf{O} \wedge d \neq \mathsf{open} \wedge eng=\mathsf{stopped}$ \\
    (ST, \mathsf{stop}, \bot, \bot) & if $\ nt=\mathsf{O} \wedge sw=1 \wedge eng \neq \mathsf{stopped}$ \\
    (ST, \bot, \bot, \bot) & if $\ nt=\mathsf{O} \wedge sw=1 \wedge eng=\mathsf{stopped}$ \\
    (f, \mathsf{up}, \bot, \bot) & if $\ nt=\mathsf{O} \wedge ds=0 \wedge ws=0 \wedge sw=0 \wedge d=\mathsf{closed} \wedge fc > f \wedge a=0$ \\
    (f, \mathsf{down}, \bot, \bot) & if $\ nt=\mathsf{O} \wedge ds=0 \wedge ws=0 \wedge sw=0 \wedge d=\mathsf{closed} \wedge fc < f \wedge a=0$ \\
    (f, \mathsf{up}, \bot, stopalarm) & if $\ nt=\mathsf{O} \wedge ds=0 \wedge ws=0 \wedge sw=0 \wedge d=\mathsf{closed} \wedge fc > f \wedge a=1$ \\
    (f, \mathsf{down}, \bot, stopalarm) & if $\ nt=\mathsf{O} \wedge ds=0 \wedge ws=0 \wedge sw=0 \wedge d=\mathsf{closed} \wedge fc < f \wedge a=1$ \\
    (f, \bot, \bot, \bot) & if $\ nt=\mathsf{O} \wedge d=\mathsf{open} \wedge sw=0$ \\
    (f, \bot, closedoor, \bot) & if $\ nt=\mathsf{O} \wedge d=\mathsf{open} \wedge fc \neq \bot \wedge a=1$ \\
    (f, \bot, \bot, \mathsf{firealarm}) & if $\ nt=\mathsf{A} \wedge sw=0$ \\
    (ST, \bot, \bot, \mathsf{firealarm}) & if $\ nt=\mathsf{A} \wedge sw=1$
\end{numcases}

\vspace{0.1cm}
$ta((f, fc, eng, d, (ws, ds), sw, a, f(at, dt1, dt2, gft, ot))) = min(at, dt1, dt2, gft, ot)$ 

\hrulefill
\end{scriptsize}
\caption{DEVS Model of an elevator control system (part c)}
\label{ElevatorDEVS3}
\end{figure}

A state, $s \in S$, of the model is a tuple that represents, respectively, the floor where elevator is, the floor where it must go, the state of the elevator engine, the state of its door, the door and weight sensors, the alarm, the display, the different timers (door and alarm), including an operational timer, artificially added to control or manage some external events, and finally, a variable indicating the next timeout.

The input, can be a number (indicating a floor) or different signals, indicating that the elevator has reached a floor, the weight or door sensor has been activated or deactivated, the opening or closing door button is pressed or the switch is turned on or off.

The output, in turn, is a tuple, where each variable represent an indication, respectively, for the display, the engine, the door and the alarm. $\bot$ means ``do nothing''.

Regarding the internal transition function let us describe some cases in order to understand it. For instance, the case (15) represents when the timer of the alarm finish and the alarm must fire, with its corresponding cases (24) and (25) in the output function case. The case (9), in turn, represents when the open button is pressed and the door must be opened, being the door closing. The corresponding case in the output function is (15).

On the other hand, the external transition function is more intuitive to understand each case of it. For instance, case (1) represents when the elevator is called from some floor, different from the actual, the engine is stopped and there is no other floor called. Meanwhile, cases (8) and (9) stand for the case when the switch is activated being the engine stopped or not respectively.

\subsubsection{Generating Simulations}
Starting with the set of all possible simulations for this example, we now apply each criteria presented before generating different classes. Later, these classes are conjoined, generating new classes. For each criterion, here, we show only some classes, the rest of them can be observed in \ref{ApElev}.

\paragraph{Transition Function Defined by Cases}
The first criterion that we apply to this example uses the definition of the external and the internal transition functions generating one class for each case in each function definition.

To describe the SCCs for this example, we will use several times a generic $s \in S$ defined as $s = (f, fc, eng, d, (ws, ds), sw, a, f(at, dt1, dt2, gft, ot),nt)$.

Some classes generated by this criterion:
\begin{footnotesize}
\begin{itemize}
    \item $IniSt_1 = \{s: s \in S \mid eng=\mathsf{stopped} \wedge fc=\emptyset\}$, \\
          $InPairs_1 = \{(x,t) \mid x=n, n \in \mathbb{N} \wedge n \neq f \wedge t \in \mathbb{R}^+_0\}$
    \item $IniSt_{13} = \{s: s \in S \mid d=\mathsf{open} \wedge fc \neq \emptyset \wedge fc \neq f\}$, \\
          $InPairs_{13} = \{(\mathsf{s_{off}},t) \mid t \in \mathbb{R}^+_0 \}$
    \item $IniSt_{30} = \{s: s \in S \mid nt=\mathsf{D_2} \wedge ds=0 \wedge ws=0 \wedge sw=0 \wedge fc > f\}$ \\
          $InPairs_{30} = \{(\tau,0)\}$
\end{itemize}
\end{footnotesize}

\paragraph{Extensional Sets}
In this example we have six sets defined by extension, $Engine$, $Door$, $Sensors$, $Alarm$, $Switch$ and $NextTimer$. Furthermore, $X$ is the union of an infinite set and a set defined by extension.

Since these sets have a relative small number of elements (considering only the finite set of the union resulting in $X$), we should define one SCC for each element of them, as this criterion proposes:
\begin{footnotesize}
\begin{itemize}
    \item $IniSt_{36} = \{s: s \in S\}$, \\
          $InPairs_{36} = \{(x,t) \mid x \in \mathbb{N}, t \in \mathbb{R}^+_0\}$
    \item $IniSt_{49} = \{s: s \in S \mid d=\mathsf{open}\}$, \\
          $InPairs_{49} = \{(x,t) \mid x \in X \cup \{\tau\}, t \in \mathbb{R}^+_0\}$
    \item $IniSt_{57} = \{s: s \in S \mid a=1\}$, \\
          $InPairs_{57} = \{(x,t) \mid x \in X \cup \{\tau\}, t \in \mathbb{R}^+_0\}$
    \item $IniSt_{59} = \{s: s \in S \mid sw=1\}$, \\
          $InPairs_{59} = \{(x,t) \mid x \in X \cup \{\tau\}, t \in \mathbb{R}^+_0\}$
\end{itemize}
\end{footnotesize}

\paragraph{Standard Partitions}
Now we apply the standard partitions criterion over the operators $<$, $>$, $+$ and $-$. For these operators could be used the same standard before in subsection \ref{StanPart}. However, some cases must be ignored since the involved variables can not be less than zero ($f \geq 0$ and $fc \geq 0$).

The following classes are some of the result of applying this criterion. The first two relate to the occurrence of the operators $<$ and $>$ in the definition of the internal transition function and the last two, to the occurrence of the operators $+$ and $-$ in the external transition function.
\begin{footnotesize}
\begin{itemize}
    \item $IniSt_{67} = \{s: s \in S \mid nt=\mathsf{O} \wedge ds=0 \wedge ws=0 \wedge sw=0 \wedge d=\mathsf{closed} \wedge fc\neq \emptyset \wedge fc=f \wedge f=0\}$, \\
          $InPairs_{67} = \{(\tau,0)\}$
    \item $IniSt_{72} = \{s: s \in S \mid nt=\mathsf{O} \wedge ds=0 \wedge ws=0 \wedge sw=0 \wedge d=\mathsf{closed} \wedge fc\neq \emptyset \wedge fc>f \wedge f=0\}$, \\
          $InPairs_{72} = \{(\tau,0)\}$
    \item $IniSt_{76} = \{s: s \in S \mid nt=\mathsf{D_2} \wedge ds=0 \wedge ws=0 \wedge sw=0 \wedge fc>f \wedge f>0\}$, \\
          $InPairs_{76} = \{(\tau,0)\}$
    \item $IniSt_{77} = \{s: s \in S \mid nt=\mathsf{D_2} \wedge ds=0 \wedge ws=0 \wedge sw=0 \wedge fc<f \wedge fc=0\}$, \\
          $InPairs_{77} = \{(\tau,0)\}$
\end{itemize}
\end{footnotesize}

\paragraph{Time Partitions}
Finally, we apply this particular criterion, taking into account the relation between the elapsed time, $e$, and the variables used for the timers: $at$, $dt1$, $dt2$, $gft$ and $ot$. Considering the values that this variables assume, the relevant key time intervals are: $[0,\mathsf{T_{D_1}}]$, $[0,\mathsf{T_{D_2}}]$, $[0,\mathsf{T_A}]$, $[0,\mathsf{T_{GF}}]$, $[\mathsf{T_{D_1}},\mathsf{T_{D_2}}]$, $[\mathsf{T_{D_1}},\mathsf{T_A}]$, $[\mathsf{T_{D_1}},\mathsf{T_{GF}}]$, $[\mathsf{T_{D_2}},\mathsf{T_A}]$, $[\mathsf{T_{D_2}},\mathsf{T_{GF}}]$, $[\mathsf{T_A},\mathsf{T_{GF}}]$ (assuming $\mathsf{T_{D_1}} < \mathsf{T_{D_2}} < \mathsf{T_A} < \mathsf{T_{GF}}$). Therefore, the relevant times $t$ to simulate the each input event are: $t=0$, $0<t<\mathsf{T_{D_1}}$, $t=\mathsf{T_{D_1}}$, $\mathsf{T_{D_1}}<t<\mathsf{T_{D_2}}$, $t=\mathsf{T_{D_2}}$, $\mathsf{T_{D_2}}<t<\mathsf{T_A}$, $t=\mathsf{T_A}$, $\mathsf{T_A}<t<\mathsf{T_{GF}}$, $T=\mathsf{T_{GF}}$ and $T>\mathsf{T_{GF}}$.
Hence, some of the classes generated by this criterion are:
\begin{footnotesize}
\begin{itemize}
    \item $IniSt_{82} = \{s: s \in S\}$ \\
          $InPairs_{82} = \{(x,t) \mid x \in X \cup \{\tau\}, t \in \mathbb{R}^+_0 \wedge \mathsf{T_{D_1}}<t<\mathsf{T_{D_2}}\}$
    \item $IniSt_{85} = \{s: s \in S\}$ \\
          $InPairs_{85} = \{(x,\mathsf{T_A}) \mid x \in X \cup \{\tau\}\}$
    \item $IniSt_{87} = \{s: s \in S\}$ \\
          $InPairs_{87} = \{(x,\mathsf{T_{GF}}) \mid x \in X \cup \{\tau\}\}$
\end{itemize}
\end{footnotesize}

\paragraph{Combining Classes}
Once all criteria are applied we make the intersection of the resulting classes obtaining new and refined ones. Here, we show only some of the classes yielded by these combinations. For example, if we want to test when someone call the elevator from some floor when the engine is stopped with the door open we must conjoin $SCC_{1}$ and $SCC_{49}$ resulting as follows:
\begin{footnotesize}
\begin{itemize}
    \newcounter{scc}
    \setcounter{scc}{89}
    \item $SCC_{1} \cap SCC_{49}:$
	\begin{align*}
		IniSt_{\arabic{scc}} &= IniSt_{1} \cap IniSt_{49} = \\
			&= \{s: s \in S \mid eng=\mathsf{stopped} \wedge fc=\emptyset\} \cap \{s: s \in S \mid d=\mathsf{open}\} = \\
			&= \{s: s \in S \mid eng=\mathsf{stopped} \wedge fc= \emptyset \wedge d=\mathsf{open}\}, \\
	\end{align*}
	\begin{align*}
		InPairs_{\arabic{scc}} &= InPairs_{1} \cap InPairs_{49} = \\
		&= \{(x,t) \mid x=n, n \in \mathbb{N} \wedge n \neq f \wedge t \in \mathbb{R}^+_0\} \cap \{(x,t) \mid x \in X \cup \{\tau\}, t \in \mathbb{R}^+_0\} = \\
		&= \{(x,t) \mid x=n, n \in \mathbb{N} \wedge n \neq f \wedge t \in \mathbb{R}^+_0\}
	\end{align*}
\end{itemize}
\end{footnotesize}
Here it can be seen the effect of the commutativity of intersection, since $SCC_{1} \cap SCC_{49}$ than $SCC_{49} \cap SCC_{1}$

Another interesting combination results from $SCC_{1}$ and $SCC_{57}$, because now, the alarm is fired:
\begin{footnotesize}
\begin{itemize}
    \addtocounter{scc}{1}
    \item $SCC_{1} \cap SCC_{57}:$
    
          $IniSt_{\arabic{scc}} = \{s: s \in S \mid eng=\mathsf{stopped} \wedge fc=\emptyset \wedge a=1\}$, \\
          $InPairs_{\arabic{scc}} = \{(x,t) \mid x \in \mathbb{N} \wedge t \in \mathbb{R}^+_0\}$
\end{itemize}
\end{footnotesize}

With the following two  classes, errors could be found if the tie-breaking rules are not well defined or they are not given at all. These are obtained by combining $SCC_{36}$ with $SCC_{87}$, on one side, and $SCC_{13}$, $SCC_{59}$ and $SCC_{85}$ on the other. For instance, the SCC defined by $IniSt_{91}$ and $InPairs_{91}$ simulates the case when the elevator is called at the same time when the ``ground floor timer'' finish.
\begin{footnotesize}
\begin{itemize}
    \addtocounter{scc}{1}
    \item $SCC_{36} \cap SCC_{87}:$
    
          $IniSt_{\arabic{scc}} = \{s: s \in S \}$, \\
          $InPairs_{\arabic{scc}} = \{(x,\mathsf{T_{GF}}) \mid x \in \mathbb{N} \}$
    \addtocounter{scc}{1}
    \item $SCC_{13} \cap SCC_{59} \cap SCC_{85}:$
    
          $IniSt_{\arabic{scc}} = \{s: s \in S \mid d=\mathsf{open} \wedge fc \neq \emptyset \wedge fc \neq f \wedge  sw=1\}$, \\
          $InPairs_{\arabic{scc}} = \{(\mathsf{s_{off}},\mathsf{T_A}) \}$
\end{itemize}
\end{footnotesize}

\subsection{Soda can vending machine}

This system consists in the control of a soda can vending machine. The machine accepts coins of \$ 0.25, \$ 0.50 and \$ 1. It gives change, optimizing it (i.e. giving the less coins as possible).

The machine has cans of two different prices (normal and diet), and the system that controls the machine must comply with the following requirements:

\begin{itemize}
\item During an operation, if after $T_{ret}$ units of time no coin is introduced into the machine or no soda is selected, the machine returns all the money that has been introduced.
\item Prices of sodas increase as time passes. Every $T_{incr}$ units of time both prices are increased in \$ 0.25.
\item if the returned money is not collected by the user after $T_{chg}$ units of time the machine recovers it.
\item The machine has a display that shows the amount of money introduced or the change after an operation.
\item At any time, before selecting a soda, the user can cancel the operation and the machine returns the money.
\end{itemize}

Some additional temporal requirements (in particular the second one) were artificially included in order to have more time variables interacting in the model allowing to increase the partitions obtained by applying the criteria of the previous section.

\subsubsection{DEVS Model}
In Figure \ref{SodaDEVS} is described a possible DEVS model for this example.
\begin{figure}
\hrulefill

\sffamily
\begin{scriptsize}
$M_{sv}$ = $(S,X,Y,\delta_{int}, \delta_{ext}, \lambda, ta)$

\vspace{0.3cm}
$S$ = $MachState \times Display \times OpTime \times NormalPrice \times DietPrice \times IncrTime \times MoneyStorage \times OperationMoney \times MoneyReturned $

\hspace{.6cm}where:

\hspace{.6cm}$MachState$ = $\{\mathsf{idle}, \mathsf{operating}, \mathsf{finishOp}, \mathsf{cancelOp}, \mathsf{waitRetChange}\}$

\hspace{.6cm}$OpTime$ = $\mathbb{R}_0^+ \cup \{\infty\}$

\hspace{.6cm}$Display$ = $IncrTime$ = $NormalPrice$ = $DietPrice$ = $\mathbb{R}_0^+$

\hspace{.6cm}$MoneyStorage$ = $OperationMoney$ = $MoneyReturned$ = $Coins1d \times Coins50c \times Coins25c$

\hspace{1cm}where:

\hspace{1cm}$Coins1d$ = $Coins50c$ = $Coins25c$ = $\mathbb{N}_0$

\vspace{0.3cm}
$X$ = $\{25, 50, 100, \mathsf{getNormal}, \mathsf{getDiet}, \mathsf{cancel}, \mathsf{moneyRetreated}\}$

\vspace{0.3cm}
$Y$ = $Display \times MoneyReturned$

\vspace{0.3cm}
$\delta_{ext}((m, d, ot, np, dp, it, ms, om, mr), e, x) = $
\setcounter{equation}{0}
\begin{numcases}{}
    (\mathsf{operating}, d + x, 0, np, dp, it - e, ms, om \oplus x, \bar{0}) & if $\ x \in \{100, 50, 25\} \wedge m \in \{\mathsf{idle}, \mathsf{operating}\}$ \\
    (\mathsf{finishOp}, d - np, 0, np, dp, it - e, ms \oplus om, \bar{0}, (d-np)\oslash(ms \oplus om)) & if $\ x = \mathsf{getNormal} \wedge d \geq np$ \\
    (\mathsf{finishOp}, d - dp, 0, np, dp, it - e, ms \oplus om, \bar{0}, \bar{0}) & if $\ x = \mathsf{getDiet} \wedge d \geq dp$ \\
    (\mathsf{cancelOp}, d, 0, np, dp, it - e, ms, \bar{0}, om) & if $\ x = \mathsf{cancel}$ \\
    (\mathsf{idle}, 0, 0, np, dp, it - e, ms, \bar{0}, \bar{0}) & if $\ x = \mathsf{moneyRetreated}$
\end{numcases}

\vspace{0.3cm}
$\delta_{int}((m, d, ot, np, dp, it, ms, om, mr)) = $
\setcounter{equation}{0}
\begin{numcases}{}
    (\mathsf{operating}, d, \mathsf{T_{ret}}, np, dp, it-ot, ms, om, \bar{0}) & if $\ m=\mathsf{operating} \wedge ot < it$ \\
    (\mathsf{waitRetChange}, d, \mathsf{T_{chg}}, np, dp, it-ot, ms \ominus (d \oslash ms), om, d \oslash ms) & if $\ m=\mathsf{finishOp} \wedge ot < it$ \\
    (\mathsf{waitRetChange}, d, \mathsf{T_{chg}}, np, dp, it-ot, ms, \bar{0}, mr) & if $\ m=\mathsf{cancelOp} \wedge ot < it$ \\
    (\mathsf{idle}, 0, \infty, np, dp, it-ot, ms \oplus mr, \bar{0}, \bar{0}) & if $\ m=\mathsf{waitRetChange} \wedge ot < it$ \\
    (\mathsf{idle}, 0, \infty, np, dp, it-ot, ms, \bar{0}, \bar{0}) & if $\ m=\mathsf{idle} \wedge ot < it$ \\
    (m, d, ot-it, np + 0.25, dp + 0.25, \mathsf{T_{incr}}, ms, om, mr) & if $\ it \leq ot$
\end{numcases}

\vspace{0.3cm}
$\lambda((m,d,ot,np,dp,it,ms,om,mr)) = (d, ms)$

\vspace{0.3cm}
$ta((m,d,ot,np,dp,it,ms,om,mr)) = min(ot, it)$

\vspace{0.3cm}
$ (coins1d, coins50c, coins25c) \oplus x =$

\hspace*{1cm}$\begin{cases}
    (coins1d + x, coins50c, coins25c) & \text{if } x = 100 \\
    (coins1d, coins50c + x, coins25c) & \text{if } x = 50 \\
    (coins1d, coins50c, coins25c + x) & \text{if } x = 25 \\
\end{cases}$

\vspace{0.3cm}
$ (coins1d, coins50c, coins25c) \oplus (coins1d', coins50c', coins25c') =$

\hspace*{1cm} $(coins1d+coins1d', coins50c+coins50c', coins25c+coins25c')$ 

\vspace{0.3cm}
$ (coins1d, coins50c, coins25c) \ominus (coins1d', coins50c', coins25c') =$

\hspace*{1cm} $(coins1d-coins1d', coins50c-coins50c', coins25c-coins25c')$

\vspace{0.3cm}
$d \oslash (coins1d, coins50c, coins25c) =
    (coins1d', coins50c', coins25c')$,

\hspace{.6cm}where:

\hspace{.6cm}$coins1d$' = $min(coins1d, d \div 1)$

\hspace{.6cm}$coins50c$' = $min(coins50c, (d - coins1d') \div 0.50)$

\hspace{.6cm}$coins25c$' = $min(coins25c, (d - coins1d' - coins50c') \div 0.25)$

\vspace{0.3cm}
$\bar{0} = (0, 0, 0)$

\hrulefill
\end{scriptsize}
\caption{DEVS Model of a soda can vending machine}
\label{SodaDEVS}
\end{figure}

In this model, a state $s \in S$ is a tuple where each variable represents, respectively, the machine state, the display, an internal timer, the actual prices (normal and diet), the timer controlling the prices increment, the money stored in the machine, the money inserted for the current operation and the change.

The input values represent, each coin denomination, the request of a normal or diet soda, the cancellation of the current operation and the signal of the change retreated.

The external transition function has one case for each input different from a coin (2, 3, 4 and 5) and one case (1) for all coins. Meanwhile, the internal transition function has one case for each internal state of the machine ($m \in MachState$) for the ``operational timer'' (1, 2, 3, 4 and 5) and one case (6) for the timeout of the ``increase price timer''.

The output consists of an ordered pair indicating what to show in the display and how much money (if any) must be returned.

\subsubsection{Generating Simulations}
As in the previous example, we start with the set of all possible simulations, apply the criteria generating different classes and later conjoin this classes obtaining new ones. As in the previous example, here we show only a few classes. The whole description of the classes obtained can be found in the \ref{ApSoda}.

\paragraph{Transition Function Defined by Cases}
Here we also use a generic $s \in S$ to describe the SCCs for this example, defined as $s = (m, d, ot, np, dp, it, ms, om, mr)$.

Some of the classes generated applying this criterion are:
\begin{footnotesize}
\begin{itemize}
    \item $IniSt_{3} = \{s: s \in S \mid d \geq dp\}$,\\
          $InPairs_{3} = \{(\mathsf{getDiet},t) \mid t \in \mathbb{R}^+_0\}$
    \item $IniSt_{7} = \{s: s \in S \mid m= \mathsf{finishOp} \wedge ot < it\}$, \\
          $InPairs_{7} = \{(\tau, 0)\}$
    \item $IniSt_{11} = \{s: s \in S \mid m= \mathsf{idle} \wedge ot \leq it\}$, \\
          $InPairs_{11} = \{(\tau, 0)\}$
\end{itemize}
\end{footnotesize}

\paragraph{Standard Partitions}
We now apply the standard partitions criterion over the partitions over the operators $\geq$, $+$, $-$, $\oplus$, $\ominus$ and $\oslash$.

\begin{itemize}
    \item $\boldsymbol{\geq}$ appears twice ($d \geq np$ and $d \geq dp$) and the standard partition for this operator is equal to the standard partition for the $<$ described in section \ref{StanPart}. The following classes are some the result of applying this criterion. The first two correspond to the application of the criterion over the operation $d \geq np$ and the last one over the operation $d \geq dp$:
    \begin{footnotesize}
    \begin{itemize}
        \item $IniSt_{12} = \{s: s \in S \mid d = np = 0\}$,\\
              $InPairs_{12} = \{(x,t) \mid x \in X \cup \{\tau\}, t \in \mathbb{R}^+_0\}$
        \item $IniSt_{14} = \{s: s \in S \mid d = 0 \wedge np > 0\}$,\\
              $InPairs_{14} = \{(x,t) \mid x \in X \cup \{\tau\}, t \in \mathbb{R}^+_0\}$
        \item $IniSt_{19} = \{s: s \in S \mid d > 0 \wedge dp > 0\}$,\\
              $InPairs_{19} = \{(x,t) \mid x \in X \cup \{\tau\}, t \in \mathbb{R}+_0\}$
    \end{itemize}
    \end{footnotesize}
    
    \item $\boldsymbol{\oplus}$, by definition, is based on $+$ and with the same type involved, $\mathbb{N}_0$, therefore they could have the same standard partition. However, since they involve only elements in $\mathbb{N}_0$ no further significant partitions can be proposed. Except if we want to simulate those cases where the implementation of those operations in the modeling language could rise some errors, e.g. overflow errors. In this case, the errors are not properly in the model but in its implementation. This is more related to a testing problem rather than validating through simulations.
    
    \item $\boldsymbol{-}$, where the operator ``$-$'' interacts with those variables used for representation of the time, the classes for those cases will be described later (Time Partitions). Some of the classes for the remaining occurrences of the operator ``$-$'' are:
    \begin{footnotesize}
    \begin{itemize}
        \item $IniSt_{22} = \{s: s \in S \mid 0 < d < np\}$, \\
              $InPairs_{22} = \{(x, t) : x \in X \cup \{\tau\}, t \in \mathbb{R}^+_0\}$
        \item $IniSt_{26} = \{s: s \in S \mid 0 < d < dp\}$, \\
              $InPairs_{26} = \{(x, t) : x \in X \cup \{\tau\}, t \in \mathbb{R}^+_0\}$
        \item $IniSt_{27} = \{s: s \in S \mid 0 < dp < d\}$, \\
              $InPairs_{27} = \{(x, t) : x \in X \cup \{\tau\}, t \in \mathbb{R}^+_0\}$
    \end{itemize}
    \end{footnotesize}
    
    \item $\boldsymbol{\oslash}$, with the operator $\oslash$ the Domain Propagation criterion can be applied  since $\oslash$ is formed by two simpler operators. Despite $-$ and $\div$ have the same standard partition, these operators involves different variables. Therefore, new classes are generated by applying the domain propagation.
    
    Some classes generated:
    \begin{footnotesize}
    \begin{itemize}
        \addtocounter{scc}{1}
        \item $IniSt_{28} = \{s: s \in S \mid d = 0\}$, \\
              $InPairs_{28} = \{(x, t) : x \in X, t \in \mathbb{R}^+_0\}$
        \item $IniSt_{29} = \{s: s \in S \mid 0 < d < coins1d \wedge coins25c = d - coins1d'-coins50c'=0\}$, \\
              $InPairs_{29} = \{(x, t) : x \in X, t \in \mathbb{R}^+_0\}$
        \item $IniSt_{31} = \{s: s \in S \mid 1 < d < coins1d \wedge 0.50 < d - coins1d' < coins50c\}$, \\
              $InPairs_{31} = \{(x, t) : x \in X, t \in \mathbb{R}^+_0\}$
    \end{itemize}
    \end{footnotesize}
\end{itemize}

\paragraph{Sets defined by extension}
In this example we have two sets defined by extension, $X$ and $MachState$. Since these two sets have a relative small number of elements, we define one SCC for each element of them, as this criterion proposes. Some of theses classes:
\begin{footnotesize}
\begin{itemize}
    \item $IniSt_{32} = \{s: s \in S \mid m = \mathsf{operating}\}$,\\
          $InPairs_{32} = \{(x,t) \mid x \in X \cup \{\tau\} \wedge t \in \mathbb{R}^+_0\}$
    \item $IniSt_{34} = \{s: s \in S \mid m = \mathsf{cancelOp}\}$,\\
          $InPairs_{34} = \{(x,t) \mid x \in X \cup \{\tau\} \wedge t \in \mathbb{R}^+_0\}$
    \item $IniSt_{41} = \{s: s \in S\}$,\\
          $InPairs_{41} = \{(x,t) \mid x = \mathsf{getNormal} \wedge t \in \mathbb{R}^+_0\}$
    \item $IniSt_{42} = \{s: s \in S\}$,\\
          $InPairs_{42} = \{(x,t) \mid x = \mathsf{getDiet} \wedge t \in \mathbb{R}^+_0\}$
\end{itemize}
\end{footnotesize}

\paragraph{Time Partitions}
In this example, the variables used to manage or simulate the time are $ot$ and $it$, besides the elapsed time $e$. Again, we have to consider the values that this variables assume to define the key time intervals in which it is relevant to simulate input events: $[0,it]$, $[0, ot]$, $[it,ot]$ (when $it < ot$) and $[ot,it]$ (when $ot < it$). Besides, a key time point is $t=ot=it$. Some classes:
\begin{footnotesize}
\begin{itemize}
    \item $IniSt_{46} = \{s: s \in S \mid it > 0 \}$, \\
          $InPairs_{46} = \{(x, t) : x \in X \cup \{\tau\}, t \in \mathbb{R}^+_0 \wedge 0<t<it \}$
    \item $IniSt_{53} = \{s: s \in S \mid 0 < it < ot \}$, \\
          $InPairs_{53} = \{(x, it) : x \in X \cup \{\tau\} \}$
    \item $IniSt_{59} = \{s: s \in S \mid 0 < ot < it \}$, \\
          $InPairs_{59} = \{(x, t) : x \in X \cup \{\tau\}, t \in \mathbb{R}^+_0 \wedge ot < t < it\}$
    \item $IniSt_{62} = \{s: s \in S \mid ot = it \}$, \\
          $InPairs_{62} = \{(x, t) : x \in X \cup \{\tau\}, t \in \mathbb{R}^+_0 \wedge t < ot\}$
    \item $IniSt_{64} = \{s: s \in S \mid ot = it \}$, \\
          $InPairs_{64} = \{(x, t) : x \in X \cup \{\tau\}, t \in \mathbb{R}^+_0 \wedge t > ot\}$
\end{itemize}
\end{footnotesize}

\paragraph{Combining Partitions}
Once that we have applied the partition criteria, we make the conjunctions between them and we keep those where the result is non empty.

Some classes obtained hereby:

\begin{footnotesize}
\begin{itemize}
    \setcounter{scc}{65}
    \item $SCC_{3} \cap SCC_{19}:$
    
          $IniSt_{\arabic{scc}} = \{s: s \in S \mid d \geq dp \wedge d>0 \wedge dp>0\}$, \\
          $InPairs_{\arabic{scc}} = \{(\mathsf{getDiet},t) \mid t \in \mathbb{R}^+_0\}$
    \addtocounter{scc}{1}
    \item $SCC_{22} \cap SCC_{41}:$
    
          $IniSt_{\arabic{scc}} = \{s: s \in S \mid 0 < d < np\}$, \\
          $InPairs_{\arabic{scc}} = \{(\mathsf{getNormal},t) \mid t \in \mathbb{R}^+_0\}$
    \addtocounter{scc}{1}
    \item $SCC_{27} \cap SCC_{42} \cap SCC_{62}:$
          
          $IniSt_{\arabic{scc}} = \{s: s \in S \mid 0 < dp < d \wedge it = ot\}$, \\
          $InPairs_{\arabic{scc}} = \{(\mathsf{getDiet},t) \mid t \in \mathbb{R}^+_0 \wedge t < ot\}$
\end{itemize}
\end{footnotesize}

Here, can be seen how an error could be detected with the SCC defined by $IniSt_{67}$ and $InPairs_{67}$, since that represents a case not defined, that is the case when the money inserted in the machine is less that the price of the soda requested.

\section{Conclusions and future work}
\label{Concl}

We present a family of criteria to conduct DEVS model simulations in a disciplined way and covering the most significant simulations to increase the confidence on the model. The main advantage of performing the simulations of a model as we propose is that users do not need the experience of a specialist or group of specialists, neither a domain expert, to select the simulations to validate the model. The selection of simulations is the result of following a set of formal rules over the mathematical model without having to know about the domain over which it is modeled. This decreases the possibility of overlooking some simulation configurations which could find errors in the model.

Another advantage of this work is the possibility of automating at least part of the validation process of DEVS models. An important open issue that needs to be addressed to enable automation is to develop a formal grammar for a mathematical language to describe DEVS models. The development of this grammar is part of our future research.

It should also be considered the possibility of re-using these techniques to test software derived from a DEVS model. A DEVS model could be used as a suitable form of the specification of a system to be implemented in some programming language. The simulation sequences generated from the application of the partition criteria could be used as test cases to test the implementation. Moreover, there exist simulation tools that generate code automatically. Thereby, if the model is thoroughly validated, the resulting piece of software would be correct. 

Future work concerns with the automation of the validation process by simulation. First it is necessary to define a standard language to write the mathematical description of a DEVS model. Afterward, we would design and develop the validation tool, including a parser for the standard language, the simulation generator and an automatic translator to some simulation tool. This would also include the definition of new coverage criteria for sequencing simulations.

Other lines of future research are: a) to extend the partition criteria to coupled models; and b) to adapt this validation technique to other formalisms.

\bibliographystyle{model1-num-names}

\appendix

\section{SCCs Generated for the Elevator}
\label{ApElev}
\subsection{Transition Function Defined by Cases}
\setcounter{scc}{1}
\begin{multicols}{2}
\begin{scriptsize}
\begin{itemize}
    \item $IniSt_{\arabic{scc}} = \{s: s \in S \mid eng=\mathsf{stopped} \wedge fc=\emptyset \}$, \\
          $InPairs_{\arabic{scc}} = \{(x,t) \mid x=n, n \in \mathbb{N} \wedge n \neq f \wedge t \in \mathbb{R}^+_0\}$
    \addtocounter{scc}{1}
    \item $IniSt_{\arabic{scc}} = \{s: s \in S \mid eng=\mathsf{up}\}$, \\
          $InPairs_{\arabic{scc}} = \{(\mathsf{fsig},t) \mid t \in \mathbb{R}^+_0\}$
    \addtocounter{scc}{1}
    \item $IniSt_{\arabic{scc}} = \{s: s \in S \mid eng=\mathsf{down}\}$, \\
          $InPairs_{\arabic{scc}} = \{(\mathsf{fsig},t) \mid t \in \mathbb{R}^+_0\}$
    \addtocounter{scc}{1}
    \item $IniSt_{\arabic{scc}} = \{s: s \in S \mid eng=\mathsf{stopped}\}$, \\
          $InPairs_{\arabic{scc}} = \{(\mathsf{ds_{on}},t) \mid t \in \mathbb{R}^+_0 \}$
    \addtocounter{scc}{1}
    \item $IniSt_{\arabic{scc}} = \{s: s \in S \mid eng \neq \mathsf{stopped}\}$, \\
          $InPairs_{\arabic{scc}} = \{(\mathsf{ds_{on}},t) \mid t \in \mathbb{R}^+_0 \}$
    \addtocounter{scc}{1}
    \item $IniSt_{\arabic{scc}} = \{s: s \in S \mid d=\mathsf{open} \wedge fc \neq \emptyset \wedge ws=0 \wedge sw=0\}$, \\
          $InPairs_{\arabic{scc}} = \{(\mathsf{ds_{off}},t) \mid t \in \mathbb{R}^+_0 \}$
    \addtocounter{scc}{1}
    \item $IniSt_{\arabic{scc}} = \{s: s \in S \mid d \neq \mathsf{open} \vee fc=\emptyset \vee ws=1 \vee sw=1\}$, \\
          $InPairs_{\arabic{scc}} = \{(\mathsf{ds_{off}},t) \mid t \in \mathbb{R}^+_0 \}$
    \addtocounter{scc}{1}
    \item $IniSt_{\arabic{scc}} = \{s: s \in S \mid eng=\mathsf{stopped}\}$, \\
          $InPairs_{\arabic{scc}} = \{(\mathsf{ws_{on}},t) \mid t \in \mathbb{R}^+_0 \}$
    \addtocounter{scc}{1}
    \item $IniSt_{\arabic{scc}} = \{s: s \in S \mid eng \neq \mathsf{stopped}\}$, \\
          $InPairs_{\arabic{scc}} = \{(\mathsf{ws_{on}},t) \mid t \in \mathbb{R}^+_0 \}$
    \addtocounter{scc}{1}
    \item $IniSt_{\arabic{scc}} = \{s: s \in S \mid fc \neq \emptyset \wedge d=\mathsf{open}\}$, \\
          $InPairs_{\arabic{scc}} = \{(\mathsf{ws_{off}},t) \mid t \in \mathbb{R}^+_0 \}$
    \addtocounter{scc}{1}
    \item $IniSt_{\arabic{scc}} = \{s: s \in S \mid fc=\emptyset \vee d \neq \mathsf{open}\}$, \\
          $InPairs_{\arabic{scc}} = \{(\mathsf{ws_{off}},t) \mid t \in \mathbb{R}^+_0 \}$
    \addtocounter{scc}{1}
    \item $IniSt_{\arabic{scc}} = \{s: s \in S\}$, \\
          $InPairs_{\arabic{scc}} = \{(\mathsf{s_{on}},t) \mid t \in \mathbb{R}^+_0 \}$
    \addtocounter{scc}{1}
    \item $IniSt_{\arabic{scc}} = \{s: s \in S \mid d=\mathsf{open} \wedge fc \neq \emptyset \wedge fc \neq f\}$, \\
          $InPairs_{\arabic{scc}} = \{(\mathsf{s_{off}},t) \mid t \in \mathbb{R}^+_0 \}$
    \addtocounter{scc}{1}
    \item $IniSt_{\arabic{scc}} = \{s: s \in S \mid fc=\emptyset\}$, \\
          $InPairs_{\arabic{scc}} = \{(\mathsf{s_{off}},t) \mid t \in \mathbb{R}^+_0 \}$
    \addtocounter{scc}{1}
    \item $IniSt_{\arabic{scc}} = \{s: s \in S \mid fc \neq \emptyset \wedge d=\mathsf{closed}\}$, \\
          $InPairs_{\arabic{scc}} = \{(\mathsf{s_{off}},t) \mid t \in \mathbb{R}^+_0 \}$
    \addtocounter{scc}{1}
    \item $IniSt_{\arabic{scc}} = \{s: s \in S \mid d=\mathsf{closing}\}$, \\
          $InPairs_{\arabic{scc}} = \{(\mathsf{od_{press}},t) \mid t \in \mathbb{R}^+_0 \}$
    \addtocounter{scc}{1}
    \item $IniSt_{\arabic{scc}} = \{s: s \in S \mid d=\mathsf{open} \wedge fc \neq \emptyset \wedge ds=0 \wedge ws=0 \wedge sw=0\}$, \\
          $InPairs_{\arabic{scc}} = \{(\mathsf{cd_{press}},t) \mid t \in \mathbb{R}^+_0 \}$
          
    \addtocounter{scc}{1}
    \item $IniSt_{\arabic{scc}} = \{s: s \in S \mid nt=O \wedge eng \neq \mathsf{stopped} \wedge f=fc \wedge f \neq 0\}$ \\
          $InPairs_{\arabic{scc}} = \{(\tau,0)\}$
    \addtocounter{scc}{1}
    \item $IniSt_{\arabic{scc}} = \{s: s \in S \mid nt=O \wedge eng \neq \mathsf{stopped} \wedge f=fc \wedge f=0\}$ \\
          $InPairs_{\arabic{scc}} = \{(\tau,0)\}$
    \addtocounter{scc}{1}
    \item $IniSt_{\arabic{scc}} = \{s: s \in S \mid nt=O \wedge eng \neq \mathsf{stopped} \wedge f \neq fc\}$ \\
          $InPairs_{\arabic{scc}} = \{(\tau,0)\}$
    \addtocounter{scc}{1}
    \item $IniSt_{\arabic{scc}} = \{s: s \in S \mid nt=O \wedge sw=1 \wedge eng \neq \mathsf{stopped}\}$ \\
          $InPairs_{\arabic{scc}} = \{(\tau,0)\}$
    \addtocounter{scc}{1}
    \item $IniSt_{\arabic{scc}} = \{s: s \in S \mid nt=O \wedge sw=1 \wedge eng=\mathsf{stopped}\}$ \\
          $InPairs_{\arabic{scc}} = \{(\tau,0)\}$
    \addtocounter{scc}{1}
    \item $IniSt_{\arabic{scc}} = \{s: s \in S \mid nt=O \wedge ds=0 \wedge ws=0 \wedge sw=0 \wedge d=\mathsf{closed} \wedge fc \neq \emptyset \wedge fc>f\}$ \\
          $InPairs_{\arabic{scc}} = \{(\tau,0)\}$
    \addtocounter{scc}{1}
    \item $IniSt_{\arabic{scc}} = \{s: s \in S \mid nt=O \wedge ds=0 \wedge ws=0 \wedge sw=0 \wedge d=\mathsf{closed} \wedge fc \neq \emptyset \wedge fc<f\}$ \\
          $InPairs_{\arabic{scc}} = \{(\tau,0)\}$
    \addtocounter{scc}{1}
    \item $IniSt_{\arabic{scc}} = \{s: s \in S \mid nt=O \wedge ds=0 \wedge ws=0 \wedge sw=0 \wedge d=\mathsf{open} \wedge fc \neq \emptyset\}$ \\
          $InPairs_{\arabic{scc}} = \{(\tau,0)\}$
    \addtocounter{scc}{1}
    \item $IniSt_{\arabic{scc}} = \{s: s \in S \mid nt=O \wedge ds=0 \wedge ws=0 \wedge sw=0 \wedge fc=\emptyset\}$ \\
          $InPairs_{\arabic{scc}} = \{(\tau,0)\}$
    \addtocounter{scc}{1}
    \item $IniSt_{\arabic{scc}} = \{s: s \in S \mid nt=O \wedge d=\mathsf{closing}\}$ \\
          $InPairs_{\arabic{scc}} = \{(\tau,0)\}$
    \addtocounter{scc}{1}
    \item $IniSt_{\arabic{scc}} = \{s: s \in S \mid nt=\mathsf{D_1} \wedge ds=0 \wedge ws=0 \wedge sw=0\}$ \\
          $InPairs_{\arabic{scc}} = \{(\tau,0)\}$
    \addtocounter{scc}{1}
    \item $IniSt_{\arabic{scc}} = \{s: s \in S \mid nt=\mathsf{D_1} \wedge \neg(ds=0 \wedge ws=0 \wedge sw=0)\}$ \\
          $InPairs_{\arabic{scc}} = \{(\tau,0)\}$
    \addtocounter{scc}{1}
    \item $IniSt_{\arabic{scc}} = \{s: s \in S \mid nt=\mathsf{D_2} \wedge ds=0 \wedge ws=0 \wedge sw=0 \wedge fc>f\}$ \\
          $InPairs_{\arabic{scc}} = \{(\tau,0)\}$
    \addtocounter{scc}{1}
    \item $IniSt_{\arabic{scc}} = \{s: s \in S \mid nt=\mathsf{D_2} \wedge ds=0 \wedge ws=0 \wedge sw=0 \wedge fc<f\}$ \\
          $InPairs_{\arabic{scc}} = \{(\tau,0)\}$
    \addtocounter{scc}{1}
    \item $IniSt_{\arabic{scc}} = \{s: s \in S \mid nt=\mathsf{D_2} \wedge \neg(ds=0 \wedge ws=0 \wedge sw=0)\}$ \\
          $InPairs_{\arabic{scc}} = \{(\tau,0)\}$
    \addtocounter{scc}{1}
    \item $IniSt_{\arabic{scc}} = \{s: s \in S \mid nt=\mathsf{A} \}$ \\
          $InPairs_{\arabic{scc}} = \{(\tau,0)\}$
    \addtocounter{scc}{1}
    \item $IniSt_{\arabic{scc}} = \{s: s \in S \mid nt=\mathsf{GF} \wedge f \neq 0 \wedge f \neq \bot \wedge d=\mathsf{open} \wedge ds=0 \wedge ws=0 \wedge sw=0\}$ \\
          $InPairs_{\arabic{scc}} = \{(\tau,0)\}$
    \addtocounter{scc}{1}
    \item $IniSt_{\arabic{scc}} = \{s: s \in S \mid nt=\mathsf{GF} \wedge f \neq 0 \wedge f \neq \bot \wedge d=\mathsf{open} \wedge \neg(ds=0 \wedge ws=0 \wedge sw=0)\}$ \\
          $InPairs_{\arabic{scc}} = \{(\tau,0)\}$    
\end{itemize}
\end{scriptsize}
\end{multicols}

\subsection{Extensional Sets}
\begin{multicols}{2}
\begin{scriptsize}
\begin{itemize}
    \addtocounter{scc}{1}
    \item $IniSt_{\arabic{scc}} = \{s: s \in S\}$, \\
          $InPairs_{\arabic{scc}} = \{(x,t) \mid x \in \mathbb{N} , t \in \mathbb{R}^+_0\}$
    \addtocounter{scc}{1}
    \item $IniSt_{\arabic{scc}} = \{s: s \in S\}$, \\
          $InPairs_{\arabic{scc}} = \{(\mathsf{fsig},t) \mid t \in \mathbb{R}^+_0\}$
    \addtocounter{scc}{1}
    \item $IniSt_{\arabic{scc}} = \{s: s \in S\}$, \\
          $InPairs_{\arabic{scc}} = \{(\mathsf{ws_{on}},t) \mid t \in \mathbb{R}^+_0\}$
    \addtocounter{scc}{1}
    \item $IniSt_{\arabic{scc}} = \{s: s \in S\}$, \\
          $InPairs_{\arabic{scc}} = \{(\mathsf{ws_{off}},t) \mid t \in \mathbb{R}^+_0\}$
    \addtocounter{scc}{1}
    \item $IniSt_{\arabic{scc}} = \{s: s \in S\}$, \\
          $InPairs_{\arabic{scc}} = \{(\mathsf{ds_{on}},t) \mid t \in \mathbb{R}^+_0\}$
    \addtocounter{scc}{1}
    \item $IniSt_{\arabic{scc}} = \{s: s \in S\}$, \\
          $InPairs_{\arabic{scc}} = \{(\mathsf{ds_{off}},t) \mid t \in \mathbb{R}^+_0\}$
    \addtocounter{scc}{1}
    \item $IniSt_{\arabic{scc}} = \{s: s \in S\}$, \\
          $InPairs_{\arabic{scc}} = \{(\mathsf{od_{press}},t) \mid t \in \mathbb{R}^+_0\}$
    \addtocounter{scc}{1}
    \item $IniSt_{\arabic{scc}} = \{s: s \in S\}$, \\
          $InPairs_{\arabic{scc}} = \{(\mathsf{cd_{press}},t) \mid t \in \mathbb{R}^+_0\}$
    \addtocounter{scc}{1}
    \item $IniSt_{\arabic{scc}} = \{s: s \in S\}$, \\
          $InPairs_{\arabic{scc}} = \{(\mathsf{s_{on}},t) \mid t \in \mathbb{R}^+_0\}$
    \addtocounter{scc}{1}
    \item $IniSt_{\arabic{scc}} = \{s: s \in S\}$, \\
          $InPairs_{\arabic{scc}} = \{(\mathsf{s_{off}},t) \mid t \in \mathbb{R}^+_0\}$
    \addtocounter{scc}{1}
    \item $IniSt_{\arabic{scc}} = \{s: s \in S \mid eng=\mathsf{up}\}$, \\
          $InPairs_{\arabic{scc}} = \{(x,t) \mid x \in X \cup \{\tau\}, t \in \mathbb{R}^+_0\}$
    \addtocounter{scc}{1}
    \item $IniSt_{\arabic{scc}} = \{s: s \in S \mid eng=\mathsf{down}\}$, \\
          $InPairs_{\arabic{scc}} = \{(x,t) \mid x \in X \cup \{\tau\}, t \in \mathbb{R}^+_0\}$
    \addtocounter{scc}{1}
    \item $IniSt_{\arabic{scc}} = \{s: s \in S \mid eng=\mathsf{stopped}\}$, \\
          $InPairs_{\arabic{scc}} = \{(x,t) \mid x \in X \cup \{\tau\}, t \in \mathbb{R}^+_0\}$
    \addtocounter{scc}{1}
    \item $IniSt_{\arabic{scc}} = \{s: s \in S \mid d=\mathsf{open}\}$, \\
          $InPairs_{\arabic{scc}} = \{(x,t) \mid x \in X \cup \{\tau\}, t \in \mathbb{R}^+_0\}$
    \addtocounter{scc}{1}
    \item $IniSt_{\arabic{scc}} = \{s: s \in S \mid d=\mathsf{closed}\}$, \\
          $InPairs_{\arabic{scc}} = \{(x,t) \mid x \in X \cup \{\tau\}, t \in \mathbb{R}^+_0\}$
    \addtocounter{scc}{1}
    \item $IniSt_{\arabic{scc}} = \{s: s \in S \mid d=\mathsf{closing}\}$, \\
          $InPairs_{\arabic{scc}} = \{(x,t) \mid x \in X \cup \{\tau\}, t \in \mathbb{R}^+_0\}$
    \addtocounter{scc}{1}
    \item $IniSt_{\arabic{scc}} = \{s: s \in S \mid ws=0\}$, \\
          $InPairs_{\arabic{scc}} = \{(x,t) \mid x \in X \cup \{\tau\}, t \in \mathbb{R}^+_0\}$
    \addtocounter{scc}{1}
    \item $IniSt_{\arabic{scc}} = \{s: s \in S \mid ws=1\}$, \\
          $InPairs_{\arabic{scc}} = \{(x,t) \mid x \in X \cup \{\tau\}, t \in \mathbb{R}^+_0\}$
    \addtocounter{scc}{1}
    \item $IniSt_{\arabic{scc}} = \{s: s \in S \mid ds=0\}$, \\
          $InPairs_{\arabic{scc}} = \{(x,t) \mid x \in X \cup \{\tau\}, t \in \mathbb{R}^+_0\}$
    \addtocounter{scc}{1}
    \item $IniSt_{\arabic{scc}} = \{s: s \in S \mid ds=1\}$, \\
          $InPairs_{\arabic{scc}} = \{(x,t) \mid x \in X \cup \{\tau\}, t \in \mathbb{R}^+_0\}$
    \addtocounter{scc}{1}
    \item $IniSt_{\arabic{scc}} = \{s: s \in S \mid a=0\}$, \\
          $InPairs_{\arabic{scc}} = \{(x,t) \mid x \in X \cup \{\tau\}, t \in \mathbb{R}^+_0\}$
    \addtocounter{scc}{1}
    \item $IniSt_{\arabic{scc}} = \{s: s \in S \mid a=1\}$, \\
          $InPairs_{\arabic{scc}} = \{(x,t) \mid x \in X \cup \{\tau\}, t \in \mathbb{R}^+_0\}$
    \addtocounter{scc}{1}
    \item $IniSt_{\arabic{scc}} = \{s: s \in S \mid sw=0\}$, \\
          $InPairs_{\arabic{scc}} = \{(x,t) \mid x \in X \cup \{\tau\}, t \in \mathbb{R}^+_0\}$
    \addtocounter{scc}{1}
    \item $IniSt_{\arabic{scc}} = \{s: s \in S \mid sw=1\}$, \\
          $InPairs_{\arabic{scc}} = \{(x,t) \mid x \in X \cup \{\tau\}, t \in \mathbb{R}^+_0\}$
    \addtocounter{scc}{1}
    \item $IniSt_{\arabic{scc}} = \{s: s \in S \mid fc=n \in \mathbb{N}\}$, \\
          $InPairs_{\arabic{scc}} = \{(x,t) \mid x \in X \cup \{\tau\}, t \in \mathbb{R}^+_0\}$
    \addtocounter{scc}{1}
    \item $IniSt_{\arabic{scc}} = \{s: s \in S \mid fc=\emptyset\}$, \\
          $InPairs_{\arabic{scc}} = \{(x,t) \mid x \in X \cup \{\tau\}, t \in \mathbb{R}^+_0\}$
    \addtocounter{scc}{1}
    \item $IniSt_{\arabic{scc}} = \{s: s \in S \mid nt = \mathsf{A}\}$, \\
          $InPairs_{\arabic{scc}} = \{(x,t) \mid x \in X \cup \{\tau\}, t \in \mathbb{R}^+_0\}$
    \addtocounter{scc}{1}
    \item $IniSt_{\arabic{scc}} = \{s: s \in S \mid nt = \mathsf{D_1}\}$, \\
          $InPairs_{\arabic{scc}} = \{(x,t) \mid x \in X \cup \{\tau\}, t \in \mathbb{R}^+_0\}$
    \addtocounter{scc}{1}
    \item $IniSt_{\arabic{scc}} = \{s: s \in S \mid nt = \mathsf{D_2}\}$, \\
          $InPairs_{\arabic{scc}} = \{(x,t) \mid x \in X \cup \{\tau\}, t \in \mathbb{R}^+_0\}$
    \addtocounter{scc}{1}
    \item $IniSt_{\arabic{scc}} = \{s: s \in S \mid nt = \mathsf{GF}\}$, \\
          $InPairs_{\arabic{scc}} = \{(x,t) \mid x \in X \cup \{\tau\}, t \in \mathbb{R}^+_0\}$
    \addtocounter{scc}{1}
    \item $IniSt_{\arabic{scc}} = \{s: s \in S \mid nt = \mathsf{O}\}$, \\
          $InPairs_{\arabic{scc}} = \{(x,t) \mid x \in X \cup \{\tau\}, t \in \mathbb{R}^+_0\}$          
\end{itemize}
\end{scriptsize}
\end{multicols}

\subsection{Standard Partitions}
\begin{multicols}{2}
\begin{scriptsize}
\begin{itemize}
    \addtocounter{scc}{1}
    \item $IniSt_{\arabic{scc}} = \{s: s \in S \mid nt=\mathsf{O} \wedge ds=0 \wedge ws=0 \wedge sw=0 \wedge d = \mathsf{closed} \wedge fc \neq \emptyset \wedge fc=f \wedge f=0\}$, \\
          $InPairs_{\arabic{scc}} = \{(\tau,0)\}$
    \addtocounter{scc}{1}
    \item $IniSt_{\arabic{scc}} = \{s: s \in S \mid nt=\mathsf{O} \wedge ds=0 \wedge ws=0 \wedge sw=0 \wedge d = \mathsf{closed} \wedge fc \neq \emptyset \wedge fc=f \wedge f>0\}$, \\
          $InPairs_{\arabic{scc}} = \{(\tau,0)\}$
    \addtocounter{scc}{1}
    \item $IniSt_{\arabic{scc}} = \{s: s \in S \mid nt=\mathsf{O} \wedge ds=0 \wedge ws=0 \wedge sw=0 \wedge d = \mathsf{closed} \wedge fc \neq \emptyset \wedge fc<f \wedge fc>0\}$, \\
          $InPairs_{\arabic{scc}} = \{(\tau,0)\}$
    \addtocounter{scc}{1}
    \item $IniSt_{\arabic{scc}} = \{s: s \in S \mid nt=\mathsf{O} \wedge ds=0 \wedge ws=0 \wedge sw=0 \wedge d = \mathsf{closed} \wedge fc \neq \emptyset \wedge fc>f \wedge f>0\}$, \\
          $InPairs_{\arabic{scc}} = \{(\tau,0)\}$
    \addtocounter{scc}{1}
    \item $IniSt_{\arabic{scc}} = \{s: s \in S \mid nt=\mathsf{O} \wedge ds=0 \wedge ws=0 \wedge sw=0 \wedge d = \mathsf{closed} \wedge fc \neq \emptyset \wedge fc<f \wedge fc=0\}$, \\
          $InPairs_{\arabic{scc}} = \{(\tau,0)\}$
    \addtocounter{scc}{1}
    \item $IniSt_{\arabic{scc}} = \{s: s \in S \mid nt=\mathsf{O} \wedge ds=0 \wedge ws=0 \wedge sw=0 \wedge d = \mathsf{closed} \wedge fc \neq \emptyset \wedge fc>f \wedge f=0\}$, \\
          $InPairs_{\arabic{scc}} = \{(\tau,0)\}$
    \addtocounter{scc}{1}
    \item $IniSt_{\arabic{scc}} = \{s: s \in S \mid nt=\mathsf{D_2} \wedge ds=0 \wedge ws=0 \wedge sw=0 \wedge fc=f \wedge f=0\}$, \\
          $InPairs_{\arabic{scc}} = \{(\tau,0)\}$
    \addtocounter{scc}{1}
    \item $IniSt_{\arabic{scc}} = \{s: s \in S \mid nt=\mathsf{D_2} \wedge ds=0 \wedge ws=0 \wedge sw=0 \wedge fc=f \wedge f>0\}$, \\
          $InPairs_{\arabic{scc}} = \{(\tau,0)\}$
    \addtocounter{scc}{1}
    \item $IniSt_{\arabic{scc}} = \{s: s \in S \mid nt=\mathsf{D_2} \wedge ds=0 \wedge ws=0 \wedge sw=0 \wedge fc<f \wedge fc>0\}$, \\
          $InPairs_{\arabic{scc}} = \{(\tau,0)\}$
    \addtocounter{scc}{1}
    \item $IniSt_{\arabic{scc}} = \{s: s \in S \mid nt=\mathsf{D_2} \wedge ds=0 \wedge ws=0 \wedge sw=0 \wedge fc>f \wedge f>0\}$, \\
          $InPairs_{\arabic{scc}} = \{(\tau,0)\}$
    \addtocounter{scc}{1}
    \item $IniSt_{\arabic{scc}} = \{s: s \in S \mid nt=\mathsf{D_2} \wedge ds=0 \wedge ws=0 \wedge sw=0 \wedge fc<f \wedge fc=0\}$, \\
          $InPairs_{\arabic{scc}} = \{(\tau,0)\}$
    \addtocounter{scc}{1}
    \item $IniSt_{\arabic{scc}} = \{s: s \in S \mid nt=\mathsf{D_2} \wedge ds=0 \wedge ws=0 \wedge sw=0 \wedge fc>f \wedge f=0\}$, \\
          $InPairs_{\arabic{scc}} = \{(\tau,0)\}$
\end{itemize}
\end{scriptsize}
\end{multicols}

\subsection{Time Partitions}
\begin{multicols}{2}
\begin{scriptsize}
\begin{itemize}
    \addtocounter{scc}{1}
    \item $IniSt_{\arabic{scc}} = \{s: s \in S\}$ \\
          $InPairs_{\arabic{scc}} = \{(x,0) \mid x \in X \cup \{\tau\}\}$
    \addtocounter{scc}{1}
    \item $IniSt_{\arabic{scc}} = \{s: s \in S\}$ \\
          $InPairs_{\arabic{scc}} = \{(x,t) \mid x \in X \cup \{\tau\}, t \in \mathbb{R}^+_0 \wedge 0<t<\mathsf{T_{D_1}}\}$
    \addtocounter{scc}{1}
    \item $IniSt_{\arabic{scc}} = \{s: s \in S\}$ \\
          $InPairs_{\arabic{scc}} = \{(x,\mathsf{T_{D_1}}) \mid x \in X \cup \{\tau\}\}$
    \addtocounter{scc}{1}
    \item $IniSt_{\arabic{scc}} = \{s: s \in S\}$ \\
          $InPairs_{\arabic{scc}} = \{(x,t) \mid x \in X \cup \{\tau\}, t \in \mathbb{R}^+_0 \wedge \mathsf{T_{D_1}}<t<\mathsf{T_{D_2}}\}$
    \addtocounter{scc}{1}
    \item $IniSt_{\arabic{scc}} = \{s: s \in S\}$ \\
          $InPairs_{\arabic{scc}} = \{(x,\mathsf{T_{D_2}}) \mid x \in X \cup \{\tau\}\}$
    \addtocounter{scc}{1}
    \item $IniSt_{\arabic{scc}} = \{s: s \in S\}$ \\
          $InPairs_{\arabic{scc}} = \{(x,t) \mid x \in X \cup \{\tau\}, t \in \mathbb{R}^+_0 \wedge \mathsf{T_{D_2}}<t<\mathsf{T_A}\}$
    \addtocounter{scc}{1}
    \item $IniSt_{\arabic{scc}} = \{s: s \in S\}$ \\
          $InPairs_{\arabic{scc}} = \{(x,\mathsf{T_A}) \mid x \in X \cup \{\tau\}\}$
    \addtocounter{scc}{1}
    \item $IniSt_{\arabic{scc}} = \{s: s \in S\}$ \\
          $InPairs_{\arabic{scc}} = \{(x,t) \mid x \in X \cup \{\tau\}, t \in \mathbb{R}^+_0 \wedge \mathsf{T_A}<t<\mathsf{T_{GF}}\}$
    \addtocounter{scc}{1}
    \item $IniSt_{\arabic{scc}} = \{s: s \in S\}$ \\
          $InPairs_{\arabic{scc}} = \{(x,\mathsf{T_{GF}}) \mid x \in X \cup \{\tau\}\}$
    \addtocounter{scc}{1}
    \item $IniSt_{\arabic{scc}} = \{s: s \in S\}$ \\
          $InPairs_{\arabic{scc}} = \{(x,t) \mid x \in X \cup \{\tau\}, t \in \mathbb{R}^+_0 \wedge t>\mathsf{T_{GF}}\}$
\end{itemize}
\end{scriptsize}
\end{multicols}

\section{SCCs Generted for the Soda Can Vending Machine}
\label{ApSoda}
\subsection{Transition Function Defined by Cases}
\setcounter{scc}{1}
\begin{multicols}{2}
\begin{scriptsize}
\begin{itemize}
    \item $IniSt_{\arabic{scc}} = \{s: s \in S \mid m \in \{\mathsf{idle}, \mathsf{operating}\}\}$,\\
          $InPairs_{\arabic{scc}} = \{(x,t) \mid x \in \{100, 50, 25\} \wedge t \in \mathbb{R}^+_0\}$
    \addtocounter{scc}{1}
    \item $IniSt_{\arabic{scc}} = \{s: s \in S \mid d \geq np\}$,\\
          $InPairs_{\arabic{scc}} = \{(\mathsf{getNormal},t) \mid t \in \mathbb{R}^+_0\}$
    \addtocounter{scc}{1}
    \item $IniSt_{\arabic{scc}} = \{s: s \in S \mid d \geq dp\}$,\\
          $InPairs_{\arabic{scc}} = \{(\mathsf{getDiet},t) \mid t \in \mathbb{R}^+_0\}$
    \addtocounter{scc}{1}
    \item $IniSt_{\arabic{scc}} = \{s: s \in S\}$,\\
          $InPairs_{\arabic{scc}} = \{(\mathsf{cancel},t) \mid t \in \mathbb{R}^+_0\}$
    \addtocounter{scc}{1}
    \item $IniSt_{\arabic{scc}} = \{s: s \in S\}$,\\
          $InPairs_{\arabic{scc}} = \{(\mathsf{moneyRetreated},t) \mid t \in \mathbb{R}^+_0\}$
    \addtocounter{scc}{1}
    \item $IniSt_{\arabic{scc}} = \{s: s \in S \mid m= \mathsf{operating} \wedge ot < it\}$, \\
          $InPairs_{\arabic{scc}} = \{(\tau, 0)\}$
    \addtocounter{scc}{1}
    \item $IniSt_{\arabic{scc}} = \{s: s \in S \mid m= \mathsf{finishOp} \wedge ot < it\}$, \\
          $InPairs_{\arabic{scc}} = \{(\tau, 0)\}$
    \addtocounter{scc}{1}
    \item $IniSt_{\arabic{scc}} = \{s: s \in S \mid m= \mathsf{cancelOp} \wedge ot < it\}$, \\
          $InPairs_{\arabic{scc}} = \{(\tau, 0)\}$
    \addtocounter{scc}{1}
    \item $IniSt_{\arabic{scc}} = \{s: s \in S \mid m= \mathsf{waitRetChange} \wedge ot < it\}$, \\
          $InPairs_{\arabic{scc}} = \{(\tau, 0)\}$
    \addtocounter{scc}{1}
    \item $IniSt_{\arabic{scc}} = \{s: s \in S \mid m= \mathsf{idle} \wedge ot < it\}$, \\
          $InPairs_{\arabic{scc}} = \{(\tau, 0)\}$
    \addtocounter{scc}{1}
    \item $IniSt_{\arabic{scc}} = \{s: s \in S \mid m= \mathsf{idle} \wedge it \leq ot\}$, \\
          $InPairs_{\arabic{scc}} = \{(\tau, 0)\}$
\end{itemize}
\end{scriptsize}
\end{multicols}

\subsection{Standard Partitions}
\begin{multicols}{2}
\begin{scriptsize}
\begin{itemize}
    \addtocounter{scc}{1}
    \item $IniSt_{\arabic{scc}} = \{s: s \in S \mid d = np = 0\}$,\\
          $InPairs_{\arabic{scc}} = \{(x,t) \mid x \in X \cup \{\tau\}, t \in \mathbb{R}^+_0\}$
    \addtocounter{scc}{1}
    \item $IniSt_{\arabic{scc}} = \{s: s \in S \mid d > 0 \wedge np = 0\}$,\\
          $InPairs_{\arabic{scc}} = \{(x,t) \mid x \in X \cup \{\tau\}, t \in \mathbb{R}^+_0\}$
    \addtocounter{scc}{1}
    \item $IniSt_{\arabic{scc}} = \{s: s \in S \mid d = 0 \wedge np > 0\}$,\\
          $InPairs_{\arabic{scc}} = \{(x,t) \mid x \in X \cup \{\tau\}, t \in \mathbb{R}^+_0\}$
    \addtocounter{scc}{1}
    \item $IniSt_{\arabic{scc}} = \{s: s \in S \mid d > 0 \wedge np > 0\}$,\\
          $InPairs_{\arabic{scc}} = \{(x,t) \mid x \in X \cup \{\tau\}, t \in \mathbb{R}^+_0\}$
    \addtocounter{scc}{1}
    \item $IniSt_{\arabic{scc}} = \{s: s \in S \mid d = dp = 0\}$,\\
          $InPairs_{\arabic{scc}} = \{(x,t) \mid x \in X \cup \{\tau\}, t \in \mathbb{R}+_0\}$
    \addtocounter{scc}{1}
    \item $IniSt_{\arabic{scc}} = \{s: s \in S \mid d > 0 \wedge dp = 0\}$,\\
          $InPairs_{\arabic{scc}} = \{(x,t) \mid x \in X \cup \{\tau\}, t \in \mathbb{R}+_0\}$
    \addtocounter{scc}{1}
    \item $IniSt_{\arabic{scc}} = \{s: s \in S \mid d = 0 \wedge dp > 0\}$,\\
          $InPairs_{\arabic{scc}} = \{(x,t) \mid x \in X \cup \{\tau\}, t \in \mathbb{R}+_0\}$
    \addtocounter{scc}{1}
    \item $IniSt_{\arabic{scc}} = \{s: s \in S \mid d > 0 \wedge dp > 0\}$,\\
          $InPairs_{\arabic{scc}} = \{(x,t) \mid x \in X \cup \{\tau\}, t \in \mathbb{R}+_0\}$
    \addtocounter{scc}{1}
    \item $IniSt_{\arabic{scc}} = \{s: s \in S \mid d = np = 0\}$, \\
          $InPairs_{\arabic{scc}} = \{(x, t) : x \in X \cup \{\tau\}, t \in \mathbb{R}^+_0\}$
    \addtocounter{scc}{1}
    \item $IniSt_{\arabic{scc}} = \{s: s \in S \mid d = np \wedge np > 0\}$, \\
          $InPairs_{\arabic{scc}} = \{(x, t) : x \in X \cup \{\tau\}, t \in \mathbb{R}^+_0\}$
    \addtocounter{scc}{1}
    \item $IniSt_{\arabic{scc}} = \{s: s \in S \mid 0 < d < np\}$, \\
          $InPairs_{\arabic{scc}} = \{(x, t) : x \in X \cup \{\tau\}, t \in \mathbb{R}^+_0\}$
    \addtocounter{scc}{1}
    \item $IniSt_{\arabic{scc}} = \{s: s \in S \mid 0 < np < d\}$, \\
          $InPairs_{\arabic{scc}} = \{(x, t) : x \in X \cup \{\tau\}, t \in \mathbb{R}^+_0\}$
    \addtocounter{scc}{1}
    \item $IniSt_{\arabic{scc}} = \{s: s \in S \mid d = dp = 0\}$, \\
          $InPairs_{\arabic{scc}} = \{(x, t) : x \in X \cup \{\tau\}, t \in \mathbb{R}^+_0\}$
    \addtocounter{scc}{1}
    \item $IniSt_{\arabic{scc}} = \{s: s \in S \mid d = dp \wedge dp > 0\}$, \\
          $InPairs_{\arabic{scc}} = \{(x, t) : x \in X \cup \{\tau\}, t \in \mathbb{R}^+_0\}$
    \addtocounter{scc}{1}
    \item $IniSt_{\arabic{scc}} = \{s: s \in S \mid 0 < d < dp\}$, \\
          $InPairs_{\arabic{scc}} = \{(x, t) : x \in X \cup \{\tau\}, t \in \mathbb{R}^+_0\}$
    \addtocounter{scc}{1}
    \item $IniSt_{\arabic{scc}} = \{s: s \in S \mid 0 < dp < d\}$, \\
          $InPairs_{\arabic{scc}} = \{(x, t) : x \in X \cup \{\tau\}, t \in \mathbb{R}^+_0\}$
    \addtocounter{scc}{1}
    \item $IniSt_{\arabic{scc}} = \{s: s \in S \mid d = 0\}$, \\
              $InPairs_{\arabic{scc}} = \{(x, t) : x \in X, t \in \mathbb{R}^+_0\}$
    
    For the operation $d \oslash (coins1d, coins50c, coins25c)$ there exist 351 SCCs, we only show some of them:
    \addtocounter{scc}{1}
    \item $IniSt_{\arabic{scc}} = \{s: s \in S \mid 0 < d < coins1d \wedge coins25c = d - coins1d'-coins50c'=0\}$, \\
              $InPairs_{\arabic{scc}} = \{(x, t) : x \in X, t \in \mathbb{R}^+_0\}$
    \addtocounter{scc}{1}
    \item $IniSt_{\arabic{scc}} = \{s: s \in S \mid coins1d < d = 1 \wedge 0 < coins25c < d - coins1d'- coins50c'\}$, \\
              $InPairs_{\arabic{scc}} = \{(x, t) : x \in X, t \in \mathbb{R}^+_0\}$
    \addtocounter{scc}{1}
    \item $IniSt_{\arabic{scc}} = \{s: s \in S \mid 1 < d < coins1d \wedge 0.50 < d - coins1d' < coins50c\}$, \\
              $InPairs_{\arabic{scc}} = \{(x, t) : x \in X, t \in \mathbb{R}^+_0\}$
    \addtocounter{scc}{1}
    \item $IniSt_{\arabic{scc}} = \{s: s \in S \mid d > 0 \wedge coins1d = coins50c = coins25c = 0 \}$, \\
              $InPairs_{\arabic{scc}} = \{(x, t) : x \in X, t \in \mathbb{R}^+_0\}$
\end{itemize}
\end{scriptsize}
\end{multicols}

\subsection{Sets defined by extension}
\begin{multicols}{2}
\begin{scriptsize}
\begin{itemize}
    \addtocounter{scc}{1}
    \item $IniSt_{\arabic{scc}} = \{s: s \in S \mid m  = \mathsf{idle}\}$,\\
          $InPairs_{\arabic{scc}} = \{(x,t) \mid x \in X \cup \{\tau\} \wedge t \in \mathbb{R}^+_0\}$
    \addtocounter{scc}{1}
    \item $IniSt_{\arabic{scc}} = \{s: s \in S \mid m = \mathsf{operating}\}$,\\
          $InPairs_{\arabic{scc}} = \{(x,t) \mid x \in X \cup \{\tau\} \wedge t \in \mathbb{R}^+_0\}$
    \addtocounter{scc}{1}
    \item $IniSt_{\arabic{scc}} = \{s: s \in S \mid m = \mathsf{finishOp}\}$,\\
          $InPairs_{\arabic{scc}} = \{(x,t) \mid x \in X \cup \{\tau\} \wedge t \in \mathbb{R}^+_0\}$
    \addtocounter{scc}{1}
    \item $IniSt_{\arabic{scc}} = \{s: s \in S \mid m = \mathsf{cancelOp}\}$,\\
          $InPairs_{\arabic{scc}} = \{(x,t) \mid x \in X \cup \{\tau\} \wedge t \in \mathbb{R}^+_0\}$
    \addtocounter{scc}{1}
    \item $IniSt_{\arabic{scc}} = \{s: s \in S \mid m = \mathsf{waitRetChange}\}$,\\
          $InPairs_{\arabic{scc}} = \{(x,t) \mid x \in X \cup \{\tau\} \wedge t \in \mathbb{R}^+_0\}$
    \addtocounter{scc}{1}
    \item $IniSt_{\arabic{scc}} = \{s: s \in S\}$,\\
          $InPairs_{\arabic{scc}} = \{(x,t) \mid x = 25 \wedge t \in \mathbb{R}^+_0\}$
    \addtocounter{scc}{1}
    \item $IniSt_{\arabic{scc}} = \{s: s \in S\}$,\\
          $InPairs_{\arabic{scc}} = \{(x,t) \mid x = 50 \wedge t \in \mathbb{R}^+_0\}$
    \addtocounter{scc}{1}
    \item $IniSt_{\arabic{scc}} = \{s: s \in S\}$,\\
          $InPairs_{\arabic{scc}} = \{(x,t) \mid x = 100 \wedge t \in \mathbb{R}^+_0\}$
    \addtocounter{scc}{1}
    \item $IniSt_{\arabic{scc}} = \{s: s \in S\}$,\\
          $InPairs_{\arabic{scc}} = \{(x,t) \mid x = \mathsf{getNormal} \wedge t \in \mathbb{R}^+_0\}$
    \addtocounter{scc}{1}
    \item $IniSt_{\arabic{scc}} = \{s: s \in S\}$,\\
          $InPairs_{\arabic{scc}} = \{(x,t) \mid x = \mathsf{getDiet} \wedge t \in \mathbb{R}^+_0\}$
    \addtocounter{scc}{1}
    \item $IniSt_{\arabic{scc}} = \{s: s \in S\}$,\\
          $InPairs_{\arabic{scc}} = \{(x,t) \mid x = \mathsf{cancel} \wedge t \in \mathbb{R}^+_0\}$
    \addtocounter{scc}{1}
    \item $IniSt_{\arabic{scc}} = \{s: s \in S\}$,\\
          $InPairs_{\arabic{scc}} = \{(x,t) \mid x = \mathsf{moneyRetreated} \wedge t \in \mathbb{R}^+_0\}$
\end{itemize}
\end{scriptsize}
\end{multicols}

\subsection{Time Partitions}
\begin{multicols}{2}
\begin{scriptsize}
\begin{itemize}
    \addtocounter{scc}{1}
    \item $IniSt_{\arabic{scc}} = \{s: s \in S \}$, \\
          $InPairs_{\arabic{scc}} = \{(\tau, 0) \}$
    \addtocounter{scc}{1}
    \item $IniSt_{\arabic{scc}} = \{s: s \in S \mid it > 0 \}$, \\
          $InPairs_{\arabic{scc}} = \{(x, t) : x \in X \cup \{\tau\}, t \in \mathbb{R}^+_0 \wedge 0<t<it \}$
    \addtocounter{scc}{1}
    \item $IniSt_{\arabic{scc}} = \{s: s \in S \mid it > 0 \}$, \\
          $InPairs_{\arabic{scc}} = \{(x, it) : x \in X \cup \{\tau\} \}$
    \addtocounter{scc}{1}
    \item $IniSt_{\arabic{scc}} = \{s: s \in S \}$, \\
          $InPairs_{\arabic{scc}} = \{(x, t) : x \in X \cup \{\tau\}, t \in \mathbb{R}^+_0 \wedge t>it\}$
    \addtocounter{scc}{1}
    \item $IniSt_{\arabic{scc}} = \{s: s \in S \mid ot > 0 \}$, \\
          $InPairs_{\arabic{scc}} = \{(x, t) : x \in X \cup \{\tau\}, t \in \mathbb{R}^+_0 \wedge 0<t<ot \}$
    \addtocounter{scc}{1}
    \item $IniSt_{\arabic{scc}} = \{s: s \in S \mid ot > 0 \}$, \\
          $InPairs_{\arabic{scc}} = \{(x, ot) : x \in X \cup \{\tau\} \}$
    \addtocounter{scc}{1}
    \item $IniSt_{\arabic{scc}} = \{s: s \in S \}$, \\
          $InPairs_{\arabic{scc}} = \{(x, t) : x \in X \cup \{\tau\}, t \in \mathbb{R}^+_0 \wedge t>ot\}$
    \addtocounter{scc}{1}
    \item $IniSt_{\arabic{scc}} = \{s: s \in S \mid 0 < it < ot \}$, \\
          $InPairs_{\arabic{scc}} = \{(x, t) : x \in X \cup \{\tau\}, t \in \mathbb{R}^+_0 \wedge 0 < t < it\}$
    \addtocounter{scc}{1}
    \item $IniSt_{\arabic{scc}} = \{s: s \in S \mid 0 < it < ot \}$, \\
          $InPairs_{\arabic{scc}} = \{(x, it) : x \in X \cup \{\tau\} \}$
    \addtocounter{scc}{1}
    \item $IniSt_{\arabic{scc}} = \{s: s \in S \mid 0 < it < ot \}$, \\
          $InPairs_{\arabic{scc}} = \{(x, t) : x \in X \cup \{\tau\}, t \in \mathbb{R}^+_0 \wedge it < t < ot\}$
    \addtocounter{scc}{1}
    \item $IniSt_{\arabic{scc}} = \{s: s \in S \mid 0 < it < ot \}$, \\
          $InPairs_{\arabic{scc}} = \{(x, ot) : x \in X \cup \{\tau\} \}$
    \addtocounter{scc}{1}
    \item $IniSt_{\arabic{scc}} = \{s: s \in S \mid 0 < it < ot \}$, \\
          $InPairs_{\arabic{scc}} = \{(x, t) : x \in X \cup \{\tau\}, t \in \mathbb{R}^+_0 \wedge t > ot\}$
    \addtocounter{scc}{1}
    \item $IniSt_{\arabic{scc}} = \{s: s \in S \mid 0 < ot < it \}$, \\
          $InPairs_{\arabic{scc}} = \{(x, t) : x \in X \cup \{\tau\}, t \in \mathbb{R}^+_0 \wedge 0 < t < ot\}$
    \addtocounter{scc}{1}
    \item $IniSt_{\arabic{scc}} = \{s: s \in S \mid 0 < ot < it \}$, \\
          $InPairs_{\arabic{scc}} = \{(x, ot) : x \in X \cup \{\tau\} \}$
    \addtocounter{scc}{1}
    \item $IniSt_{\arabic{scc}} = \{s: s \in S \mid 0 < ot < it \}$, \\
          $InPairs_{\arabic{scc}} = \{(x, t) : x \in X \cup \{\tau\}, t \in \mathbb{R}^+_0 \wedge ot < t < it\}$
    \addtocounter{scc}{1}
    \item $IniSt_{\arabic{scc}} = \{s: s \in S \mid 0 < ot < it \}$, \\
          $InPairs_{\arabic{scc}} = \{(x, it) : x \in X \cup \{\tau\} \}$
    \addtocounter{scc}{1}
    \item $IniSt_{\arabic{scc}} = \{s: s \in S \mid 0 < ot < it \}$, \\
          $InPairs_{\arabic{scc}} = \{(x, t) : x \in X \cup \{\tau\}, t \in \mathbb{R}^+_0 \wedge t > it\}$
    \addtocounter{scc}{1}
    \item $IniSt_{\arabic{scc}} = \{s: s \in S \mid ot = it \}$, \\
          $InPairs_{\arabic{scc}} = \{(x, t) : x \in X \cup \{\tau\}, t \in \mathbb{R}^+_0 \wedge t < ot\}$
    \addtocounter{scc}{1}
    \item $IniSt_{\arabic{scc}} = \{s: s \in S \mid ot = it \}$, \\
          $InPairs_{\arabic{scc}} = \{(x, t) : x \in X \cup \{\tau\}, t \in \mathbb{R}^+_0 \wedge t = ot\}$
    \addtocounter{scc}{1}
    \item $IniSt_{\arabic{scc}} = \{s: s \in S \mid ot = it \}$, \\
          $InPairs_{\arabic{scc}} = \{(x, t) : x \in X \cup \{\tau\}, t \in \mathbb{R}^+_0 \wedge t > ot\}$
\end{itemize}
\end{scriptsize}
\end{multicols}

\end{document}